\documentclass[12pt]{article}

\usepackage[latin1]{inputenc}
\usepackage{amsmath}
\usepackage{amsfonts}
\usepackage{amssymb,slashed}
\usepackage{cite}
\usepackage{amsxtra}
\usepackage[margin=3cm]{geometry}
\usepackage{color}
\usepackage{hyperref}
\hypersetup{linktocpage=true}
\usepackage{graphicx}
\usepackage{mathrsfs}
\usepackage{multirow}
\usepackage{tikz}

\numberwithin{equation}{section}
\newcommand{\beq}{\begin{equation}}
\newcommand{\eeq}{\end{equation}}
\def\be {\begin{equation}}
\def\ee {\end{equation}}
\def\ba#1\ea{\begin{align}#1\end{align}}
\def\baed#1\eaed{\begin{aligned}#1\end{aligned}}
\def\bged#1\eged{\begin{gathered}#1\end{gathered}}
\def\bea{\begin{eqnarray}}
\def\eea{\end{eqnarray}}

\def\a{\alpha}
\def\b{\beta}

\def\e{\epsilon}

\def\F{\Phi}
\def\g{\gamma}
\def\G{\Gamma}

\def\L{\Lambda}

\def\o{\omega}
\def\O{\Omega}

\def\r{\rho}
\def\s{\sigma}

\def\z{\zeta}

\def\Re{\text{Re}}
\def\Im{\text{Im}}

\let\foo\bar
\renewcommand{\bar}[1]{ {\foo{  #1} }{} }

\newlength{\dhatheight}

\usepackage{amsmath,latexsym,amssymb,slashed}
\usepackage{bbm}
\usepackage{pdfpages}
\usepackage{centernot}
\usepackage{multirow}
\usepackage{varwidth}
\usepackage{cite}
\usepackage{amsthm}
\usepackage{xypic}

\theoremstyle{definition}
\newtheorem{defn}{Definition}[section]
\newtheorem{propn}{Proposition}[section]
\newtheorem{thms}[propn]{Theorem}

\newcommand{\eq}[1]{\begin{equation}\begin{split}#1\end{split}\end{equation}}
\newcommand{\spl}[1]{\begin{split}#1\end{split}}
\newcommand{\al}[1]{\begin{align}#1\end{align}}

\newcommand{\dfn}[1]{\begin{defn}#1\end{defn}}

\newcommand{\thm}[1]{\begin{thms}#1\end{thms}}
\newcommand{\prf}[1]{\begin{proof}#1\end{proof}}

\newcommand{\arxth}[1]{\href{http://arxiv.org/abs/hep-th/#1}{[{\tt hep-th/#1}]}}
\newcommand{\arx}[1]{[\href{http://arxiv.org/abs/#1}{\tt #1}]}

\newcommand{\bcal}{\mathcal{B}}

\newcommand{\dcal}{\mathcal{D}}

\newcommand{\fcal}{\mathcal{F}}

\newcommand{\hcal}{\mathcal{H}}
\newcommand{\ical}{\mathcal{I}}
\newcommand{\jcal}{\mathcal{J}}

\newcommand{\ncal}{\mathcal{N}}

\newcommand{\zbb}{\mathbbm{Z}}
\newcommand{\rbb}{\mathbbm{R}}
\newcommand{\cbb}{\mathbbm{C}}

\renewcommand{\t}{\theta}

\newcommand{\vt}{\vartheta}

\newcommand{\ea}{\bigwedge\nolimits^{\!\bullet} T^*}

\newcommand{\p}{\partial}

\newcommand{\vol}{\text{vol}}
\def\d{\text{d}}

\renewcommand{\Re}{\text{Re}\;}
\renewcommand{\Im}{\text{Im}~}

\numberwithin{equation}{section}

\begin{document}

\baselineskip=16pt
\setlength{\parskip}{6pt}

\begin{titlepage}
\begin{flushright}
\parbox[t]{1.4in}{
\flushright IPhT-T16/003}
\end{flushright}

\begin{center}

\vspace*{1.7cm}

{\LARGE \bf  Mirror symmetry \& supersymmetry on $SU(4)$-structure backgrounds}

\vskip 1.6cm

\renewcommand{\thefootnote}{}

\begin{center}
 \normalsize
 Ruben Minasian$^{\,a,b}  $ and Dani\"el Prins$^{\,a,c}$
\end{center}
\vskip 0.1cm
 {\sl $^a$ Institut de physique th\'eorique, Universit\'e Paris Saclay, CNRS, CEA \\F-91191 Gif-sur-Yvette,
France}
\vskip 0.1cm
{\sl $^b$ School of Physics, Korea Institute for Advanced Study, Seoul 130-722, Korea}
\vskip 0.1cm
{\sl $^c$ Dipartimento di Fisica, Universit\`{a} di Milano-Bicocca,\\
 I-20126 Milano, Italy}
\vskip 0.2 cm
{\textsf{ruben.minasian@cea.fr}} \\
{\textsf{daniel.prins@cea.fr}}
\end{center}

\vskip 1.5cm
\renewcommand{\thefootnote}{\arabic{footnote}}

\begin{center} {\bf ABSTRACT } \end{center}
We revisit the backgrounds of type IIB on manifolds with $SU(4)$-structure and discuss two sets of solutions  arising from internal geometries that are complex and symplectic respectively. Both can be realized in terms of generalized complex geometry. We identify a map which relates  the complex and symplectic supersymmetric systems. In the semi-flat torus bundle setting this map corresponds to T-duality and suggest a way of extending the mirror transform to non-K\"ahler geometries.

\end{titlepage}

\newpage
\tableofcontents
\vspace{20pt}

\setcounter{page}{1}
\setlength{\parskip}{9pt}

\section{Introduction and discussion}

Mirror symmetry relating compactifications of string theory on two very different internal geometries has proven to be extremely powerful in the study of Calabi-Yau compactifications. The SYZ approach \cite{syz} provides an intrinsic construction of the mirror pairs by asserting that every Calabi-Yau  $X$ with a mirror $\check{X}$ is a $T^3$ special Lagrangian fibration over a three-dimensional base, and the mirror symmetry is T-duality along the three circles
of the $T^3$. The mirror manifold $\check{ X}$ is the moduli space of the space of the special Lagrangian torus, making the exchange of complex and symplectic properties of $X$ and $\check{ X}$ natural. It also provides the most promising way of extending the mirror program beyond Calabi-Yau manifolds.

In spite of much recent progress in our ability to describe flux backgrounds with generally non-Ricci flat internal geometries, our understanding of mirror symmetry in a this more general setting is rather limited. While it is generally believed that  the string (flux) backgrounds should display mirror symmetry, precise definitions are generally missing and the question on how practical this mirror transform can be is still open. There is some progress in extending the SYZ approach to K\"ahler manifolds (see e.g. \cite{aur, gkr}) and generalized Calabi-Yau manifolds \cite{sng}, and the immediate  questions are which geometries admit mirrors and how to construct a mirror map.  One immediate place to look for generalizations is in the supersymmetric geometries, i.e. internal manifolds that can lead to string compactifications preserving supersymmetry. Moreover since the backgrounds are not purely geometric and involve fluxes, so should the mirror map. Hence it might be natural to try to look for mirrors in the generalized geometric context.

Generalized complex geometry provides a convenient framework for describing the supersymmetric backgrounds with six internal dimensions \cite{gmpt}. One of the necessary conditions of preservation of supersymetry is that the internal manifold admits a generalized Calabi-Yau structure, i.e. a closed pure spinor of $Spin(6,6)$. The parity of the closed pure spinor changes depending on whether one is in type IIA (even) or in IIB (odd). This means in particular that when the supersymmetry of the internal manifold is related to an $SU(3)$-structure, the supersymmetric geometries suitable for IIA compactifications are symplectic and for IIB - complex. So using some loose definition of mirror map as an exchange of complex and symplectic properties, one can declare that IIA and IIB four-dimensional compactifications are mirror to each other. Unfortunately there are no good examples of actual pairs of mirror backgrounds. And of course there is no general proof that any supersymmetric background with a mirror should have an internal space given by a torus fibration.

Working in a  semi-flat torus bundle setting and assuming that there are commuting (local) isometries is often a useful approach as this allows to construct explicit maps using the Fourier-Mukai transform. Notably one could start form a particularly simple case of the prototypical non-K\"ahler background of type IIB with a RR three-form $F_3$, which is constrained by supersymmetry to be
\eq{\label{eq:RR3}
F_3 = -i( \p - \bar{\p}) J \equiv \d^c J\;,
}
where $J$ is a positive real two-form (the fundamental form), and study its image under the Fourier-Mukai transform. Without being concerned for now about the supersymmetry equations including  \eqref{eq:RR3}, being compatible with $T^3$ fibrations one can find such an image.\footnote{For an S-dual system involving NSNS three-form only, often referred to as Strominger system \cite{str}, mirror symmetry relates two different backgrounds of the same type \cite{fmt}. The RR forms necessarily change parity under three T-dualities.}   Note that one has to take special care to stay within the class of $SU(3)$-structure solutions here (for us, T-duality has to be ``maximally type-changing" with respect to the generalized almost complex structures). One could think of several generalizations, such as the inclusion of other fluxes, in particular the NSNS three-form flux $H$.  Examples of various T-dualities in compact string backgrounds, e.g. involving nilmanifolds have been considered in literature \cite{kstt, gmpt3,kt}, though most of the T-duals one would be very reluctant to call mirrors as the number of T-dualities is less than three.

In addition to considering the non-Calabi-Yau setting, the extension of the procedure to higher dimension is another interesting direction, and it was pursued in \cite{lty}. The first step of the approach there was to write the special case
of the four-dimensional  IIA and IIB supersymmetry constraints (without NSNS-flux, but with RR-flux)  in terms of a set of geometrical equations, which we shall refer to as the {\it LTY 2A system} and {\it LTY 2B system} respectively. The LTY 2B is a more familiar system as it involves complex manifolds with balanced metrics, and can be seen as the type II analogue of the Strominger system. The LTY 2A system in the case of four (external) dimensions involves symplectic half-flat manifolds, i.e., $\d J = \d \text{Re}(\O) = 0$. Provided the manifolds of the 2A and 2B system admit a  $T^3$-fibration, it is then possible to show that the Fourier-Mukai transform relates the LTY 2B system to the LTY 2A system, and hence defines a mirror map. The second step of the construction of \cite{lty} generalizes the LTY 2A and 2B systems to manifolds of arbitrary real dimension $2n$, along with the Fourier-Mukai transform, in such a way that, given dual $T^n$-fibrations, the 2B system is still mapped to the 2A system. The LTY 2B system still requires a complex balanced manifold, but the half-flatness condition of the LTY 2A system is replaced by a more convoluted condition which depends on the fibration structure.

Although this procedure  yields a  notion of higher dimensional non-Calabi-Yau mirror symmetry, it is not clear to what extent the LTY 2A/2B systems correspond to the constraints of supersymmetry in dimensions other than four. An obstacle here is that a generalization of the simple description of supersymmetry in terms of generalized complex geometry to higher-dimensional internal manifolds has proven to be non-trivial.\footnote{In addition to four,  the fulll analysis of supersymmetry exists in six external dimensions \cite{lpt}.} In two external dimensions, the main case of interest for us, such a description exists under certain caveats, which will be reviewed shortly. In zero external dimensions, the system of polyform equations does not fully capture the supersymmetry equations \cite{tom, su5}.

We shall revisit a class of type IIB supersymmetric solutions on eight-manifolds of $SU(4)$-structure and examine how the LTY 2A and 2B systems are related to such supersymmetric solutions. We shall also re-examine the Fourier-Mukai and T-duality transformations that relate the mirror backgrounds.

One problem that one has to face in passing from six to eight internal dimensions is that generic (Weyl) spinors on the internal manifold are no longer pure. As a  consequence, repackaging the supersymmetry equations into nice pure spinor equations, while still possible, is not in general one-to-one. One can still have nice and compact equations written in terms of pure spinors which however do not capture the full content of supersymmetry conditions (and the difference is far from being pretty!). Fortunately, the situation can be ameliorated by making two assumptions; first, that the internal manifold admits an $SU(4)$-structure, and second, that the internal components of the Killing spinors are pure. In this case, solving the system \eqref{susy} is equivalent to preserving supersymmetry.
The second problem is the complexity of the internal equations, and here we shall deal with the two simplest cases. The  $SU(4)$-structure implies existence of two chiral nowhere vanishing spinors. The two simplest constructions used here assume that either the second spinor is proportional to the first, or to its complex conjugate. In fact, the first  case, labeled as the {\sl strict $SU(4)$ ansatz}, has been discussed in \cite{pt} and as we shall see here is related to the LTY 2B system. We shall refer to this case as the {\it complex supersymmetric system}. As we shall show, the second case yields a different supersymmetric system which is related to the LTY 2A system. We shall refer to this case as the {\it symplectic supersymmetric system}.

We shall briefly review the $SU(4)$-structure supersymmetry conditions (and similarities and differences with the six-dimensional case). As mentioned already, the most important fact is that for $SU(4)$-structure internal manifolds, supersymmetry is fully captured by the pure spinor equations. The pure spinors are respectively
\eq{\label{eq:prsp}
\Phi_+ &= - e^{-i \vt} e^{-\phi}e^{B- i J} \\
\Phi_- &= - e^{i \vt} e^{-\phi} e^B \O \,.}
Both $\Phi_+$ and $\Phi_-$ are even (type 0 and 4 respectively), but we borrow here  from $SU(3)$ nomenclature. Just like for six-dimensional internal spaces, the conformal closure of one of the two is a necessary condition for supersymmetry preservation. Depending on which pure spinor is closed, we shall label the resulting systems as symplectic (closed $e^{2A}\Phi_+$, where $A$ is the warp factor) or complex (closed  $e^{2A}\Phi_-$). The roles played by $\Phi_+$ and $\Phi_-$ are swapped upon going from one spinorial ansatz to the other. For both systems we consider compactifications of type IIB strings and both reductions yield two-dimensional $\ncal =(2,0)$ theories.\footnote{ For type IIA one again expects a pair of pure spinors of the same parity. Due to RR fluxes being even, the pure spinors should be odd (type 1 and 3) and a (local) vector field is needed in the construction. We shall not consider type IIA compactifications here.}

The closed pure spinor, lets call it $\Phi_2$ will define an integrable generalized almost complex structure (even more, in fact - a generalized Calabi-Yau structure), which we shall denote as $\jcal_2$. The imaginary part of the compatible pure spinor $\Phi_1$ then yields a pair of equations that determine the sources and impose further geometric constrains. These equations are more conveniently written using polyforms $\Psi_i$ ($i=1,2$) defined in \eqref{eq:poly}, ($\Phi_i = e^{-\phi} e^B \Psi_i$):
\eq{
\d_H \d_H^{\jcal_2} \left( e^{- \phi} \text{Im} \Psi_1 \right) &= \rho  \\
\d_H^{\jcal_2} \left( e^{- \phi} \text{Im} \Psi_1 \right) &= - e^{-2A}\s \star \d_H \left( e^{2A - \phi} \text{Re} \Psi_1 \right)
}
where $\d_{ H} \equiv \d + H \wedge$, $\d^\jcal \equiv [\d, \jcal]$,  and $\s$ flips some of the individual signs of $k$-forms in the expansion of a polyform, $\s \Psi_{(k)}  = (-1)^{\frac12 k (k-1)} \Psi_{(k)}$ . We have obtained the above by rewriting \eqref{susy} in order to facilitate comparison with the LTY system.  $\rho$ is a source term or current (for non-compact sourceless backgrounds, $\rho=0$). The first equation is familiar from the four dimensionsional supersymmetry conditions. The second equation is a constraint that does not appear in the four-dimensional case, where instead one has a conformal closure of of $\text{Re} \Psi_1$.

The similarity with the (internal) six-dimensional case will be strongest if one considers the complex supersymmetric system (see section \ref{cplxsusy}) and takes $H=0$, $F= F_3$ (all other RR fluxes vanishing), constant warp factor $A$, vanishing dilaton $\phi$, and $\vt = 0$, where $\vt$ is some parameter related to the precise ansatz for the Killing spinor. One will then recover
\eq{ \label{eq:bb}
\d \O &= 0 \\
\d J^{3} &=0 \\
2 i \p \bar{\p}  J   &\equiv \r \;
}
and an extra constraint
\eq{\label{eq:bbb}
\d^c J = \frac12 \star \d J^2.}
As we shall see \eqref{eq:bb} corresponds to LTY 2B system. Since the constraint \eqref{eq:bbb} is missing in six-dimensions it is not guessed when generalizing the six-dimensional equations to higher dimensions and is hence missing from LTY 2B conditions. Actually for the case under consideration, the constraint is automatically satisfied on any solution of \eqref{eq:bb}. It is however non-trivial in general. The relations between complex/symplectic supersymmetric systems and LTY 2B/2A systems will be discussed in detail in section \ref{compare}, where in particular we shall show that LTY 2B system is  a proper subset of the supersymmetry constraints with strict Killing spinor ansatz (and the story is much more complicated for the symplectic case). Very roughly the differences between the two sets of systems boil down to the missing constraints like \eqref{eq:bbb} on one hand, and to  imposing some constraints based on the necessity of a fibration structure on the other.

Once we understand the relations of supersymmetry equations to LTY systems, we can also discuss the map between the two supersymmetric systems. Here we find some variation from the map presented in \cite{lty}. The Fourier-Mukai map as discussed in \cite{lty} is constructed in such a way that it maps symplectic geometry to complex geometry and vice versa. From the point of supersymmetry, this seems to be not the most natural approach. Instead, one should consider a map that relates generalized complex geometry induced by a complex structure to generalized complex geometry induced by a symplectic structure.

In particular, the situation can best be described by the following diagram:
\eq{
\xymatrix{
\left(\O^\bullet_\bcal(M_B), \p, \bar{\p} \right)  \ar[d]^P \ar[rd]^{FT} & \\
\left(\O^\bullet_\bcal(M_B), \frac{i}{2} \d^c, \frac{i}{2} \d \right) \ar[r]^T &  \left(\O^\bullet_\bcal(M_A), \frac{i}{2} \d^\L, \frac{i}{2} \d \right)
}
}
This diagram should be understood as follows\footnote{ Note that the definition of the Fourier-Mukai transform \eqref{fm} is defined for the situation without NSNS fluxes $H$, $\check{H}$, and hence this diagram technically only holds true in that situation. However, the generalization to include these fluxes is straightforward.}. The Fourier-Mukai transform $FT$ is constructed to give an isomorphism between the following differential complexes:
\eq{
\left(\O^\bullet_\bcal(M_B), \p, \bar{\p} \right) \simeq \left(\O^\bullet_\bcal(M_A),  \frac{i}{2} \d^\L, \frac{i}{2}  \d \right) \;.
}
On the other hand, the map $T$ gives an isomorphism such that the generalized Dolbeault operator with respect to the generalized complex structures is preserved:
\eq{
(\O^\bullet_\bcal(\check{M}), \p^{\check{\jcal}}, \bar{\p}^{\check{\jcal}} ) \simeq (\O^\bullet_\bcal(M), \p^\jcal,  \bar{\p}^\jcal ) \;.
}
This map is the T-duality map as discussed in \cite{bem, bhm, cg}.
In particular, $T$ is an isomorphism in the case where we take $\check{\jcal} = \check{\jcal}_I$ to be induced by a complex structure $I$ and $\jcal = \jcal_J$ to be induced by a symplectic structure $J$. Then $\d^{\check{\jcal}_I} = \d^c$ and $\d^{\jcal_J} = \d^\L \equiv [\d, \L]$\footnote{Note that $\d^\L : \O^k(M) \rightarrow \O^{k-1}(M)$ lowers the degree of a form.}, with $\L$, the adjoint of the Lefschetz operator $J\wedge$, defined as contraction with $J^{-1}$. Furthermore, $P$ is the polarity switch operator defined in \eqref{pso}, $\bcal$ is the base of the $T^4$ fibration, and $\O^\bullet_\bcal(M) \subset \O^\bullet (M)$ is the sheaf of fiber-invariant differential forms.

The bottom line is that the T-duality map is the one which correctly maps the complex supersymmetric system to the symplectic supersymmetric system.  In fact, the T-duality map not only maps supersymmetric solutions to supersymmetric solutions, it also maps backgrounds to backgrounds, i.e., it preserves the supersymmetric integrability conditions.

The structure of the paper is as follows. In section \ref{lty}, we review the LTY 2A and 2B systems. In section \ref{tb}, we discuss the geometry of torus bundles, the particularities of which will be necessary when comparing the LTY 2A system to the symplectic supersymmetric system and when considering the details of the Fourier-Mukai and T-duality maps. A brief overview on supersymmetry is given in section \ref{supersymmetry}, where we then proceed to review the complex supersymmetric system which was found in \cite{pt} by taking a strict Killing spinor ansatz. We then discuss the symplectic supersymmetric system, which is a new solution to the supersymmetry equations, in section \ref{symp}. Next, we compare the LTY 2B system to the complex supersymmetric system and the LTY 2A system to the symplectic supersymmetric system in section \ref{compare}. We find that the 2B system is missing a number of constraints captured by the complex supersymmetric system, whereas the situation is more convoluted in the 2A/symplectic case.
We discuss mirror symmetry and T-duality in section \ref{ms}. First, we review the Fourier-Mukai map and discuss in what sense the LTY 2A and 2B systems are mirror. Then we revisit the T-duality map and demonstrate in what sense the complex and symplectic supersymmetric systems are T-dual. Finally, in section \ref{ftheo}, we consider a special case of  the complex supersymmetric system that involves conformally Calabi-Yau fourfolds and admits F-theory lift.  In appendix \ref{integ}, we review details of the integrability of the complex supersymmetric solution and give some simple examples of backgrounds. We then proceed to do the same for the symplectic supersymmetric solution. The other appendices give some technical details necessary for the computation of the supersymmetric systems.

A word on notation:
We shall use $I$ to denote almost complex structures, $J$ to denote almost symplectic structures (i.e., positive-definite real two-forms), $\O$ to denote a volume form of the canonical line bundle, and $\jcal$ to denote generalized almost complex structures. Furthermore, we denote the external dimension by $d$ and the complex internal dimension by $n$. We use the word `background' to refer to solutions of the equations of motion in $D=10$, whereas the term `supersymmetric solution' refers to solutions of the supersymmetry equations (specifically, \eqref{eq:susyeq}).

\section{The LTY systems}\label{lty}
In this section, we review the supersymmetric $SU(n)$-structures of \cite{lty}. These consist of a number of tensors $(J, \O, K, \rho)$, which are supposed to encode the geometry and RR fluxes, and a number of constraints, which should capture the requirements of supersymmetry (for some unspecified $\ncal$) in arbitrary dimension. There are two kinds of such supersymmetric $SU(n)$-structures, the 2A and 2B system, with the difference being the kind of constraints imposed. They describe symplectic and complex geometries respectively, and the two systems are mirror to one another in some fashion that will be discussed in section \ref{ms}. For $n=3$  ($d=4$), these systems are equivalent to supersymmetry preservation conditions (in absence of NSNS three-form flux $H$). The relation of these systems to preservation of supersymmetry in higher dimensions (specifically, $n=4$) will be discussed in detail in section \ref{compare}.

We shall consider $2n$-dimensional internal manifolds which admit an $SU(n)$-structure.

\dfn{An {\it $SU(n)$-structure} $(J, \O)$ on a manifold $M$ of real dimension $2n$ consists of a real positive-definite two-form $J$, the {\it almost symplectic structure}, and a complex nowhere vanishing decomposable $n$-form $\O$, satisfying
\eq{\label{eq:su}
J \wedge \O &= 0 \\
\frac{1}{2^n} \O \wedge \O^* &= \frac{i^n}{n!} K J^n = \text{vol}_n
}
The nowhere vanishing function $K$ is known as the conformal factor of the $SU(n)$-structure.
$\O$ determines an almost complex structure with respect to which  $\O=\O^{(4,0)}$, $J= J^{(1,1)}$.
}
Calabi-Yau manifolds are particular cases of $SU(n)$-structure spaces, which require the additional constraint $\d J = \d \O = 0$  and $K$ constant. Generically, a manifold carrying an $SU(n)$-structure need neither be complex nor symplectic. A proposal to geometrize the type IIB supersymmetry constraints on an $SU(n)$-structure manifold was given in \cite{lty}.

\dfn{A {\it supersymmetric $SU(n)$-structure of type 2B} on $M$ is given by an $SU(n)$-structure $(J_B, \O_B)$ satisfying
\eq{ \label{y2b}
\frac{1}{2^n} \O_B \wedge \O_B^* &= \frac{i^n}{n!} K_B J_B^n \\
\d J^{n-1}_B &=0 \\
\d \O_B &= 0 \\
2i  \p \bar{\p} \left( K_B^{-1} J_B \right) &\equiv \r_B \;.
}
In particular, $(M,J_B, \O_B)$ is complex and balanced. We will also refer to \eqref{y2b} as the {\it LTY 2B system}.
}
In the case of $n=3$, these equations reduce to the polyform equations of \cite{gmpt} with $H=0$ imposed. These polyform equations determine solutions to the $\ncal = 1$ supersymmetry equations of type IIB on $\rbb^{1,3} \times M_6$. From this point of view, $\rho_B$ determines a source for the RR Bianchi identities. However, note that the equation involving $\rho_B$ is not so much a restriction on the system as it is a definition of $\rho_B$ as long as we do not provide an explicit definition of $\rho_B$ in terms of the RR fluxes.

Similar to the LTY 2B system, there is a 2A system that determines solutions to the $\ncal =1 $ supersymmetry equations of type IIA on $\rbb^{1,3} \times M_6$.
It turns out that, in order to ensure that the generalization to arbitrary dimensions has certain desired properties, we will need to introduce the following.

\dfn{Let $M$ be equipped with an $SU(n)$-structure. Let $U \subset M$ be open and dense. Then a {\it special real polarization} $\dcal$ is an integrable distribution that is 1) Lagrangian with respect to $J$, and 2) special with respect to $\O$. In other words, $\forall p \in U$, $\dcal_p \subset T_p M$ is a dimension $n$ subspace that satisfies $J|_{\dcal_p} = 0$ and $\O|_{\dcal_p} \in e^{i \varphi} \rbb^+$ for some phase $\varphi$. }

The notion that $U$ should be dense but need not be $M$ has to do with the possibility of degenerations which we will specify more clearly later on.
We will not take such degenerations into account. Therefore, for all intents and purposes, we will consider $U=M$.

Given a special real polarization and a metric (or a connection) the tangent bundle splits as $T = \dcal \oplus \dcal^\perp$. Hence this induces a decomposition of the exterior algebra and differential forms,
\eq{
\O^k(M) = \bigoplus_{a+b = k} \O^{(a,b)^\dcal}(M) \;,
}
analogous to how the presence of an almost complex structure decomposes $k$-forms into $(p,q)$-forms. In particular, an $(a,b)$-form has $a$ $\dcal^*$ directions (or `legs'). By making use of the associated projection operators, we define the following.
\dfn{A {\it supersymmetric $SU(n)$-structure of type 2A} on $M$ is given by an $SU(n)$-structure $(J_A, \O_A)$ satisfying
\eq{ \label{y2a}
\frac{1}{2^n}\O_A \wedge \O_A^* &= \frac{i^n}{n!} K_A J_A^n \\
\d J_A &=0 \\
\d \left( \pi_\dcal^{0,n} + \pi_\dcal^{n-1,1}\right) \O_A &= 0 \\
- i \d \d^\dcal \left( K_A \left( \pi_\dcal^{1,n-1} + \pi_\dcal^{n,0}\right) \O_A \right) &\equiv \r_A
}
In particular, $(M,J_A, \O_A)$ is symplectic.\footnote{Note that we have switched the definition of all projection operators from $\pi^{a,b}_\dcal$ to $\pi^{b,a}_\dcal$ with respect to the definitions of \cite{lty}. As  in section 3 of \cite{lty}, we define $\dcal^*$ as spanned by $\{\d r^1,...,\d r^n\}$. Then $\pi^{k, n-k}_\dcal$  maps onto forms with $k$ directions $\d r^j$.}
}
For a certain phase in the definition of the real polarization, the projection operators are such that for $n=3$, $\left(\pi_\dcal^{0,n} + \pi_\dcal^{n-1,1}\right) \O_A = \Re \O_A$ while $\left( \pi_\dcal^{n-1,1} + \pi_\dcal^{n,0}\right) \O_A = \Im \O_A$, thus reducing these equations to the polyform equations of \cite{gmpt} for type IIA supersymmetry. On the other hand, for $n=4$, instead one finds that
\eq{
\Re \O_A &= (\pi_\dcal^{4,0}+ \pi_\dcal^{2,2} + \pi_\dcal^{0,4} ) \O_A \\
\Im \O_A &= (\pi_\dcal^{3,1} + \pi_\dcal^{1,3}) \O_A \;.
}
We will return to this decomposition in section \ref{2asymp}.

\section{Torus fibrations}\label{tb}
In order to discuss the mirror relation between  the 2A and 2B systems, it will be necessary to restrict the geometry to the case of torus bundles. In particular, the natural habitat of the 2A system is a Lagrangian fibration, which comes with an induced special real polarization. The 2B system then naturally lives on the dual torus fibration, which is complex. The picture one should keep in mind is as follows:
$$
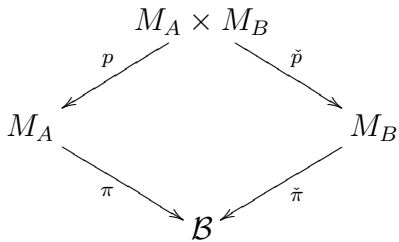
\xymatrix{
& M_A \times M_B \ar[ld]_{p} \ar[rd]^{\check{p}} & \\
M_A \ar[rd]_{\pi} & & M_B\ar[ld]^{\check{\pi}} \\
&\bcal&
}
$$
where the {\it correspondence space} $M_A \times M_B$ is defined as
\eq{
M_A \times M_B = \{ (x,y) \in (M_A, M_B) \; | \; \pi(x) = \check{\pi}(y) \in \bcal \} \;.
}
Here, $M_A$ is symplectic whereas $M_B$ is complex; both are of real dimension $2n$, with $\bcal$ of real dimension $n$. Note, however, that when we are just comparing the supersymmetric solutions to the 2A and 2B system in section \ref{compare}, it is only the 2A system for which this fibration structure plays a significant role. The fibration structure on the 2B side will only be relevant when considering the actual mirror mapping in section \ref{ms}.

In this section, we will discuss the geometrical aspects of torus fibrations which can be equipped with $SU(n)$-structures. Given a torus bundle with local coordinates $\{r^1, ..., r^n\}$ on the base and $\{\t^1, ..., \t^n\}$ on the fibers, we define the following:

\dfn{A {\it fiber-invariant form} $\a$ is a form that can locally be written as
\eq{
\a = \a_{j_1...j_a k_1...k_b} (r) \d r^{j_1} \wedge ... \wedge \d r^{j_a} \wedge \d \t^{k_1} \wedge ... \wedge \d \t^{k_b} \;.
}
Specifically, the restriction is that the coefficients do not depend on the coordinates $\t^j$.
The sheaf of all fiber-invariant forms on a fibration $M \rightarrow \bcal$ is denoted by $\O^\bullet_\bcal(M)$.}

\dfn{Let  $M \rightarrow \bcal$ be a Lagrangian torus bundle with $SU(n)$-structure $(J, \O)$. The $SU(n)$-structure is said to be {\it semi-flat} if $J, \O \in \O_\bcal^\bullet(M)$.}

We will restrict ourselves to such semi-flat $SU(n)$-structures. Furthermore, we will also neglect singular fibers. We follow mostly along the lines of \cite{gross}, section 5, which we adapt here to discuss the not-necessarily Calabi-Yau case.

\subsection{Lagrangian fibrations}
Consider the symplectic manifold $(M_A, J_A)$, which is a fibration $\pi : M_A \rightarrow \bcal$ over the base space $\bcal$ with Lagrangian toroidal fibers. By the Arnold-Liouville theorem, locally there exist action-angle coordinates $(\t^j, r^j)$ such that $r^1, ..., r^n$ are coordinates on the base and $\t^1, ...,\t^n$ are coordinates on the toroidal fibers. The action-angle coordinates define a tropical affine structure. That is to say,
under a change of coordinates $(r^j, \t^j) \rightarrow (\tilde{r}^j, \tilde{\t}^j)$, one has that
\eq{\label{eq:coords}
r^j = B^j_{\phantom{j}k} \tilde{r}^k + c^j  \;,
}
with $B \in GL(n, \zbb)$, $c \in \rbb^n$.
As a consequence, we can conclude that we can define an integrable distribution $\dcal$ locally spanned by $\{\p_{r^1}, ... , \p_{r^n}\}$.
In order to ensure existence of a complex decomposable $n$-form $\O_A$, we assume that $B$ defined in \eqref{eq:coords} satisfies $B \in SL(n, \zbb)$. Hence Lagrangian torus fibrations satisfying this assumption are special cases of $SU(n)$-structure manifolds with a polarization. By choosing $\O_A$ correctly, the polarization is real, hence such a space satisfies the necessary requirements to admit an LTY 2A system.

Let us investigate the properties of the $SU(n)$-structure. First, note that the $SU(n)$-structure defines an almost complex structure. Locally, we can define a frame of holomorphic one-forms as
\eq{
\tilde{\zeta}^j \equiv \tilde{A}^j_{\phantom{j}k}(r)\d \t^k + \tilde{B}^j_{\phantom{j}k}(r)\d r^k \;,
}
with $j \in \{1,...,n\}$, and $\tilde{A}, \tilde{B}$ invertible; they are fiber-invariant since the almost complex structure is fiber-invariant. As the almost complex structure is invariant under $GL(n, \cbb)$, we can change frame to
\eq{\label{eq:acs}
\zeta^j &\equiv \d \t^j +  \b^j_{\phantom{j}k} \d r^k \;, ~~~~~~~~\qquad \b \equiv \tilde{A}^{-1}\tilde{B}\\
 &\equiv \d \t^j + A^j_{\phantom{j}k} \d r^k + i \r^j_{\phantom{j}k} \d r^k \;.
}
Here, $A \in \text{End}(n, \rbb)$, $\r \in GL(n, \rbb)$ are the real and imaginary components of $\b$.
Since $\zeta^j$ are holomorphic, $\b$ encodes the almost complex structure. The local frame
\eq{
\delta \t^j &\equiv \d \t^j + A^j_{\phantom{j}k} \d r^k = \frac12 (\z^j + \bar{\z}^j )\\
\r^j &\equiv \r^j_{\phantom{j}k} \d r^k = \frac{1}{2i} (\zeta^j - \bar{\z}^j)
}
is such that the almost complex structure $I$ acts as
\eq{
I( \delta \t^j ) &= - \r^j  \\
I ( \r^j) &= \delta \t^j \;.
}
Since $\r^j_{\phantom{j}k}$ is invertible, locally the special real polarization is given by
\eq{\label{rp1}
\dcal^* \equiv \text{span}\{ \d r^1, ..., \d r^n \} = \text{span}\{\r^1, ..., \r^n \} \;,
}
and we define
\eq{\label{rp2}
\dcal^{* \perp} = \text{span}\{ \delta \t^1, ..., \delta \t^n \} \;,
}
such that $T^* = \dcal^* \oplus \dcal^{* \perp}$. Thus, given both $\dcal^*$ and $\dcal^{*\perp}$, it is now possible to make sense of the projection operators used in \eqref{y2a}.

We will return to the interpretation of $A, \r$ momentarily, but first, let us discuss $(J_A, \O_A)$.
We start with the properties of the symplectic form $J_A$. By definition of the action-angle coordinates, the fact that $J_A$ is a (1,1)-form with respect to the almost complex structure and the fact that $J_A$ is real and positive-definite, we find the following (see \cite{gross}):
\begin{itemize}
\item First of all, $J$ can be written in terms of various coordinate systems as
\eq{\label{sympj}
J_A &= \d \t^j \wedge \d r^j \\
&= \frac{i}{2} h_{jk} \z^j  \wedge \bar{\z}^k \\
&= h_{jk} \d \t^j \wedge \r^k \\
&= h_{jk} \delta \t^j \wedge \r^k \;.
}
\item By equating the first two, it follows that $h$ is real and symmetric and satisfies
\eq{
h_{jk}\r^{k}_{\phantom{k}l} = \delta_{jl} \;.
}
In other words, $\r = h^{-1}$. As a consequence, $\r$ is also symmetric.
\item Viewed as a matrix, $A$ is also symmetric. However, perhaps a more intuitive way to view $A$ is as a (Lie-algebra valued) connection one-form, $A \in \O^1(\bcal; \mathfrak{t})$, where the Lie algebra $\mathfrak{t}$ is isomorphic to the tangent space of the fibers of $M_A$. Thus, the real part of the integrability constraint \eqref{lfinteg} is the demand that the connection is flat: $\d A^j = 0$ (or equivalently,  $\d \delta \t^j = 0$).
\end{itemize}
Next, let us consider $\O_A$.
As $\O_A$ is a decomposable (4,0)-form, it follows that
\eq{\label{sympo}
\O_A = S(r) \zeta^1 \wedge \zeta^2 \wedge \zeta^3 \wedge \zeta^4 \;,
}
It can be shown that (see \cite{gross} Thm 5.7) in the semi-flat case, integrability of the almost complex structure is equivalent to
\eq{\label{lfinteg}
\p_{[j} \b^l_{\phantom{l}k]} = 0 \;.
}
Due to \eqref{sympj}, it follows that
\eq{
\frac{1}{2^n} i^n (-1)^{\frac12 n (n-1)} \O_A \wedge \O_A^* = \frac{1}{n!}\left|S \sqrt{\text{det}(\r)}\right|^2   J_A^n \;.
}
Thus, if we are interested in having trivial conformal factor, we should take
\eq{\label{sympconf}
S^{-1} = e^{ i (\vt + \check{\vt})}  \sqrt{\text{det}(\r)} \;,
}
with the notation for the angle chosen for later convenience.

\subsection{Complex torus fibrations}
Given a Lagrangian torus fibration $M_A$, locally one has the isomorphism  that for $U \subset \bcal$, $\pi^{-1}(U) \simeq T^*U / \dcal^*$, with $\dcal^*$ defined in \eqref{rp1}. On the other hand, one can construct another fibration $M_B$ over $\bcal$, with as fibers of $M_B$ the duals of the toroidal fibers of $M_A$. We shall refer to $M_B$ as the mirror of $M_A$. For the mirror, one has that for $U \subset \bcal$, $\check{\pi}^{-1}(U) \simeq TU / \dcal$.
Given a semi-flat $SU(n)$-structure on $M_A$, there is a canonical decomposable $n$-form $\O_B$ on $M_B$ associated to $J_A$, which determines a complex structure. Furthermore, there is no obstruction to the existence of a real positive-definite $(1,1)$-form $J_B$ such that $(J_B, \O_B)$ form an $SU(n)$-structure on $M_B$. We denote the fiber coordinates on $M_B$ by $\{ \check{\t}^1, ..., \check{\t}^n \}$. The complex coordinates $\{z^j\}$ are then such that
\eq{\label{cplxcoords}
\d z^j = \d \check{\t}^j + i \d r^j \;,
}
and
\eq{\label{cplxo}
\O_B = \check{S} e^{- i (\vt + \check{\vt})} \d z^1 \wedge ... \wedge \d z^n \;.
}
Here, $\check{S}$ is a nowhere-vanishing real function on $\bcal$, which for our purposes will be determined by supersymmetry, and we have chosen a convenient notation for the phase.
We can write the almost symplectic structure as
\eq{\label{cplxj}
J_B = \frac{i}{2} \check{h}_{jk} \d z^j \wedge \d z^k \;,
}
with $\check{h}_{jk} = \check{h}_{kj}$. From this, we conclude that the conformal factor $K$ of the $SU(n)$-structure is trivial when
\eq{\label{cplxconf}
\sqrt{\text{det}(\check{h})} = \check{S}\;.
}

\section{Supersymmetry}\label{supersymmetry}
In order to check to what extent the LTY 2A and 2B systems correspond to solutions to the supersymmetry equations in various dimensions, we examine the case of $d=2$, $\ncal= (2,0)$ with $SU(4)$-structure. As it turns out, the LTY 2B system corresponds in some sense to a known IIB solution described in \cite{pt}, which we will refer to as the {\it complex supersymmetric solution}. The LTY 2A on the other hand, corresponds to a previously unknown solution which we have found. It is, unlike what one may have expected, another IIB solution, which we shall refer to as the {\it symplectic supersymmetric solution}.

In this section, we shall describe our setup to solve the supersymmetry equations of type II supergravity, review the reformulation of supersymmetry in terms of polyform equations in the language of generalized complex geometry, and review the complex supersymmetric solution. We then proceed to discuss the new symplectic supersymmetric solution.

In order to find (bosonic) flux backgrounds of type II supergravity, the most convenient way to proceed is to insist that all fields are trivial under supersymmetry transformations, and then check what additional conditions must be met for such fields to satisfy the equations of motions. The vanishing of the supersymmetry transformations of the fermions then leads to the supersymmetry solutions, which in our conventions are given by
\eq{\label{eq:susyeq}
0 &= \left(\underline{\p}\phi + \frac{1}{2} \underline{H} \right) \epsilon_1+ \left( \frac{1}{16} e^{\phi} \G^M \underline{\fcal}\G_M \G_{11} \right) \epsilon_2  \\
0 &= \left( \underline{\p }\phi - \frac{1}{2} \underline{H} \right) \epsilon_2 - \left( \frac{1}{16} e^{\phi} \G^M \sigma(\underline{\fcal})\G_M \G_{11} \right) \epsilon_1  \\
0 &= \left( \nabla_M + \frac{1}{4} \underline{H}_M \right) \epsilon_1 + \left( \frac{1}{16} e^{\phi} \underline{\fcal}\G_M \G_{11} \right) \epsilon_2  \\
0 &= \left( \nabla_M - \frac{1}{4} \underline{H}_M \right) \epsilon_2 - \left( \frac{1}{16} e^{\phi} \sigma(\underline{\fcal})\G_M \G_{11} \right) \epsilon_1  \;.
}
Here, underlining is defined by contraction with gamma matrices, $\e_{1,2}$ are the Killing spinors, $\phi$ is the dilaton, $H$ the NSNS flux, $\fcal$ the total RR flux. We work in the democratic formalism, where the RR flux, given by
\eq{
\fcal = \sum_k \fcal_{(k)} \;,\quad k \in
 \left\{
\begin{array}{cc}
\{0,2,4,6,8,10\} & \text{IIA}\\
\{1,3,5,7,9\} & \text{IIB}
\end{array} \right.
}
needs to be supplemented with the additional selfduality constraint
\eq{\label{sd}
\fcal = \star_{10} \s \fcal \;,
}
with $\s \fcal_{(k)} = (-1)^{\frac12 k (k-1)} \fcal_{(k)} $.
Let us now reduce the most general case to the case we are interested in. First, we are interested in the case where $d=2$, i.e., $M_{10} = \rbb^{1,1} \times M_8$ with warped metric
\eq{
ds^2 (M_{10})  = e^{2 A} ds^2(\rbb^{1,1}) + ds^2(M_8) \;,
}
with $A \in C^\infty(M_8, \rbb)$ the warp factor and $ds^2(\rbb^{1,1})$ the two-dimensional Minkowski metric. As a consequence, Poincar\'{e} invariance combined with the selfduality constraint \eqref{sd} allows us to decompose the fluxes as
\eq{\label{fluxdecomp}
\fcal = \vol_2 \wedge e^{2A} \star_8 \s F + F \;.
}
A further simplifcation follows from specifying the supersymmetry. We are interested in the case $\ncal = (2,0)$, i.e., there are two positive chirality (external) supercharges, which can be denoted by a complex valued Weyl spinor $\zeta$. For such solutions the most general decomposition of the Killing spinors is given by
\eq{
\epsilon_1 &= \zeta \otimes \eta_1 + \text{c.c.} \\
\epsilon_2 &= \zeta \otimes \eta_2 + \text{c.c.} \;,
}
with $\eta_{1,2}$ Weyl spinors of $Spin(8)$ with equivalent norms and with the chirality of $\eta_j$ the same as that of $\epsilon_j$. We assume that both are nowhere vanshing. Furthermore, we assume that $M_8$ can be equipped with an $SU(4)$-structure, which is equivalent to the existence of a pure spinor $\eta$. Given such a spinor $\eta$,  transitivity of the Clifford algebra ensures that pointwise, any spinor can be expressed as
\eq{
\xi_+ &= a \eta + b \eta^c + c_{mn} \g^{mn} \eta^c \\
\xi_- &= d_m \g^m \eta + e_m \g^m \eta^c \;,
}
as can be verified by a quick dof counting (in fact this is true regardless of whether or not $\eta$ is pure). For simplicity, we will restrict our attention to the case where $\eta_{1,2}$ are also pure; this is equivalent to demanding they satisfy\footnote{See the appendix of \cite{pt} for our spinor conventions (or in fact most other conventions).}
\eq{
\tilde{\eta}_1 \eta_1 = \tilde{\eta}_2 \eta_2 = 0 \;.
}
With this setup, it has been shown \cite{pt, rosa} that the supersymmetry equations \eqref{eq:susyeq} can be recast in the framework of generalized complex geometry as
\eq{\label{susy}
\d_H \left( e^{2A - \phi} \text{Re} \Psi_1 \right) &= e^{2A} \star \s F \\
\d_H \left( e^{2 A - \phi} \Psi_2 \right) &= 0 \\
\d_H^{\jcal_2} \left( e^{- \phi} \text{Im} \Psi_1 \right) &= F \;.
}
These equations are both necessary and sufficient. This recasting is analogous to the $\rbb^{1,3} \times M_6$ scenario discussed in \cite{gmpt}. The main difference is that on $M_6$, any Weyl spinor is pure. On $M_8$, purity of a Weyl spinor is in fact an additional constraint, and there is a topological obstruction to the existence of pure spinors\footnote{An equation involving $\d^\jcal$ can be used in $d=4$ to describe supersymmetry, see \cite{tomasiello}.}.

The polyforms $\Psi_{1,2}$ are defined as
\eq{\label{eq:poly}
\Psi_1 &= -\frac{2^4}{|\eta_1|^2} \eta_1 \otimes \tilde{\eta}^c_2 \\
\Psi_2 &= -\frac{2^4}{|\eta_1|^2} \eta_1 \otimes \tilde{\eta}_2
}
and can be considered as spinors of $Spin(6,6)$. By purity of $\eta_{1,2}$, $\Psi_{1,2}$ are also pure and hence correspond to generalized almost complex structures.

It should be stressed that solutions to the supersymmetry equations are not automatically solutions to the equations of motions. In fact, for this to be the case, two additional constraints need to be satisfied. Firstly, the Bianchi identities
\eq{\label{bianchi}
\d_H \fcal = \d H = 0
}
should be satisfied. Secondly, the external part of the equations of motions for $B$ should be checked by hand. This second constraints is particular to $d=2$; in $d=4$, this is automatic by demanding Poincar\'{e} invariance of the background.

\subsection{Complex supersymmetric system}\label{cplxsusy}
Let us restrict our attention to type IIB. Without loss of generality we can consider $\eta$ unimodular. Given such a unimodular pure spinor, one can construct an $SU(4)$-structure by taking
\eq{\label{su4spin}
J_{mn} &= -i \tilde{\eta}^c \g_{mn} \eta \\
\O_{mnpq} &= \tilde{\eta} \g_{mnpq} \eta \;.
}
Then $(J, \O)$ satisfies \eqref{eq:su} with $K=1$. In order to construct supersymmetry solutions, let us fix $\eta_1 = \a \eta$ without loss of generality. The {\it strict $SU(4)$-ansatz} now entails taking $\eta_2 \sim \eta_1$ and hence, since their norms ought to be equivalent,
\eq{\label{strict}
\eta_2 = \a e^{i \vt} \eta\;,
}
with $\a, \vt \in C^\infty(M_8, \rbb)$. This has been worked out in \cite{pt}. The polyforms are given by
\eq{\label{cgacs}
\Psi_1 &= - e^{-i \vt} e^{- i J} \\
\Psi_2 &= - e^{i \vt} \O \;,
}
and hence, the generalized almost complex structures are of type (0,4). After decomposing the fluxes into $SU(4)$-irreps (see appendix \ref{su4decomp}), the solution to the supersymmetry equations is then given by
\eq{ \label{pt}
W_1 &= W_2 = 0 \\
W_3 &= i e^\phi (\cos\vt f^{(2,1)}_3  - i \sin\vt f^{(2,1)}_5) \\
W_{4} &= \frac{2}{3}\p  (\phi- A )\\
W_{5} &=  \p (\phi - 2 A +  i \vt)\\
\a &= e^{ \frac{1}{2} A}\\
\tilde{f}^{(1,0)}_{3} &=  \tilde{f}^{(1,0)}_{5} = \tilde{h}^{(1,0)}_{3} =0 \\
h_{1}^{(1,0)} &= 0\\
h_{3}^{(1,0)} &= \frac{2}{3} \p \vt \\
f_{1}^{(1,0)} &=-i \p ( e^{- \phi}\sin\vt )\\
f_{3}^{(1,0)} &= - \frac{i}{3}e^{2A} \p ( e^{-2A- \phi}\cos\vt )\\
f_{5}^{(1,0)} &=\frac{1}{3}e^{-4A} \p( e^{4A- \phi}\sin\vt )\\
f_{7}^{(1,0)} &=e^{-2A} \p( e^{2A- \phi}\cos\vt )\\
h^{(2,1)} &= e^\phi (- \cos\vt f^{(2,1)}_5  + i \sin\vt f^{(2,1)}_3)
~.
}
The free parameters of the solution are the warp factor, dilaton and an internal Killing spinor parameter, $A, \phi, \vt \in C^\infty(M_8, \rbb)$, as well as the primitive parts of the RR flux $f_{3,5}^{(2,1)} \in \O^{(2,1)}(M_8)$.
Note in particular that the solution enforces $M_8$ to be complex. For this reason, we shall refer to this solution as the {\it complex susy system}.

\subsection{Symplectic supersymmetric system}\label{symp}
In this rather brief section we present a new solution to the supersymmetry equations, which is one of the main results of this paper. It is similar to the complex supersymmetric system of the previous section, in that it is a solution for IIB, $d=2$, $\ncal = (2,0)$, with an $SU(4)$-structure on $M_8$. The entire setup is the same, except for the ansatz for the internal part of the Killing spinor. Instead of considering the strict case \eqref{strict}, we take the ansatz
\eq{\label{eq:sympansatz}
\eta_2 = \a e^{- i \vartheta} \eta^c \;.
}
Note that obviously, $\eta_2$ is pure.
With this Killing spinor, one finds that
\eq{\label{sgacs}
\Psi_1 &= - e^{i \vt} \O \\
\Psi_2 &= - e^{-i \vt} e^{- i J} \;.
}
As one can see, the result is interchanging of the polyforms, and hence, the associated generalized almost complex structures are of type (4,0) instead. Solving the supersymmetry equations (either in the guise of \eqref{eq:susyeq} or \eqref{susy}) for this case leads to the following supersymmetric solution:
\eq{\label{eq:sympsusy}
W_1 &= W_3 = W_4 = 0 \\
W_2 &=   -2 i e^{\phi - i \vt} f_3^{(2,1)}\\
W_{5m} &= \p_m A = \frac12 \p_m \phi = 2 \p_m \log \a =  2 e^{\phi + i \vt} \tilde{f}_{3|m}^{(1,0)}\\
F_1 &= F_5 = F_7 = H= 0 \\
f_{3|m}^{(1,0)}  &= 0  \\
\p_m \vt &= 0
}
Since any such solution is automatically symplectic, we will refer to \eqref{eq:sympsusy} as the {\it symplectic susy system}.

\section{Comparison of $d=2$ supersymmetry with the LTY systems}\label{compare}
The LTY systems were inspired by considering solutions to supersymmetry for $d=4$ ($n=3$)\footnote{Specifically, they describe the so called $SU(3)$ case described by \cite{gmpt}, with type (0,3) and type (3,0) generalized almost complex structures.}, and then generalized to arbitrary (even) dimension to investigate mirror symmetry. However, it is not clear from this point of view whether or not the LTY 2A and 2B systems actually have anything to do with supersymmetric solutions in dimensions other than four. We will consider the case of $d=2$ on manifolds with $SU(4)$-structure. In this case, as it turns out, both the LTY 2A and 2B systems are related to supersymmetric solutions of type IIB, rather than 2A to IIA and 2B to IIB as was the case in $d=4$.\footnote{We shall nevertheless keep the ``LTY 2A" and ``LTY 2B" nomenclature for the respective systems, and hopefully this will not lead to confusion.} The reason for this is that for $SU(4)$, both $e^{-iJ}$ and $\O$ are even polyforms (or in the language of $Spin(2n,2n)$, both have the same chirality), which is contrary to the case $n=3$. Another way to view this is to note that mirror symmetry should be something akin to T-duality, and for $n=4$, there is an even number of T-dualities, thus a IIB solution gets mapped to a IIB solution, rather than a IIA solution.

\subsection{LTY 2B system $\implies$ complex supersymmetric system}\label{2bcplx}
We find that solutions to the complex supersymmetric system \eqref{pt} are solutions to the LTY 2B system \eqref{y2b}, but the converse is not true. Equivalently, at the level of the constraints, the LTY 2B constraints are a proper subset of the supersymmetry constraints with strict Killing spinor ansatz.

Let $(M_8, F, H, \phi, g)$ satisfy \eqref{pt}, with $SU(4)$-structure given by $(J, \O)$. In order to prove our claim, we must construct $(J_B, \O_B, K_B, \rho_B)$ that satisfy \eqref{y2b}. These are given by\footnote{Technically, the most general solution is
\eq{
\O_B &= k_1 e^{-\phi + 2A + i \vt + i k_3} \O\\
J_B &= k_2^{-1} e^{\frac{-i}{4}k_4} e^{\frac23 (-\phi + A)} J\\
K_B &= k_1^2 k_2^4 e^{i k_4} e^{ \frac23 (\phi+ 2 A)}\;,
}
with $k_{1,2,3,4} \in \rbb$. We do not expect these constants to be significant in any way and hence take them to be trivial.}
\eq{
\O_B &= e^{-\phi + 2A + i \vt} \O\\
J_B &= e^{\frac23 (-\phi + A)} J\\
K_B &= e^{ \frac23 (\phi+ 2 A)}\;.
}
Note that, as mentioned before, the last equation of \eqref{y2b} can be taken as definition of $\rho_B$. Verifying that $(J_B, \O_B, K_B)$ indeed satsifies \eqref{y2b} is a simple matter of plugging in the torsion classes of $(J, \O)$ and noting the shifts caused by this rescaling.

The reason why this does not fully capture supersymmetry is two-fold. Even though  NSNS flux is neglected in \cite{lty}, non-trivial $H$ is in fact not incompatible with the LTY 2B system. However,  it is not obvious how to add it by hand to \eqref{y2b}.

Secondly,  $\rho_B$ cannot fully capture the RR fluxes. Roughly speaking,
the second equation of \eqref{y2b} gives a constraint for $W_4$, the third for $W_{1,2,5}$, and the fourth equation gives a constraint for $W_3$ as well as an additional constraint for $W_4$, and separately, the vanishing of $\tilde{f}_{3,5}^{(1,0)}$. If one imagines also demanding $H=0$, then there are three constraints missing on $(1,0)$-fluxes.

In heretoric strings for $n=3$,  the equation $i \p \bar{\p} J = \frac{\a'}{4} (\rm{tr}R^2 - \rm{tr} F^2)$ imposes stringent local and global constraints on the solution.\footnote{Formally, supersymmetry conditions for $n=3$ Strominger system are of the form \eqref{y2b}.} In type II theories, its importance is also great, as ensuring that it corresponds to negative tension sources is often the only way around the no-go theorems for the compact flux backgrounds.  Note that in general (differently from the $\d H = 0$ case), $\rho_B$ is not exact. The knowledge of how to explicitly relate $\rho_B$ to RR fluxes is however important for the construction of string backgrounds.


\subsection{LTY 2A system compared to the symplectic supersymmetric system} \label{2asymp}
The case of the LTY 2A system is somewhat similar in the sense that, roughly speaking, solutions to the symplectic supersymmetric system \eqref{eq:sympsusy} should be related to solutions to the LTY system 2A system. However, this case is extremely more convoluted due to the necessity of introducing the real polarization. Clearly, the existence of a real polarization is necessary for being able to define the projection operators $\pi_\dcal^{a,b}$, and hence the LTY 2A system. The 2B case did not have to rely on existence of any such external factors; any supersymmetry solution simply induces a LTY 2B solution. Furthermore, there is an additional constraint on the complex structure for the LTY 2A system, which is not present in the symplectic supersymmetric system. We will demonstrate the precise relationship that holds for the 2A solution, and leave the explanation for this latter constraint to section \ref{tduality}.

The precise relation between the LTY 2A system and the symplectic susy system is as follows:
\thm{
Let $M_8 \rightarrow \bcal_4 $ be a Lagrangian torus bundle. Let $(J, \O)$, as determined by the pure spinor $\eta$ in \eqref{su4spin}, be a semi-flat $SU(4)$-structure.  Let $(M_8, F, H, \phi, g)$ satisfy \eqref{eq:sympsusy}. Let the almost complex structure, defined in terms of $\b$ introduced in \eqref{eq:acs}, satisfy
\eq{\label{eq:argh}
\p_{[j} A^l_{\phantom{l}k]} = 0\;,
}
i.e., the connection is flat.
Then there exists $(J_A, \O_A, K_A, \rho_A)$ satisfying \eqref{y2a}.
}
\prf{
Let us first note that $\rho_A$ is again whatever it is by definition, that $J$ is symplectic and hence we can take $J_A = J$, and that $K_A$ is fixed as soon as one has defined both $J_A$ and $\O_A$. So the problem is reduced to constructing $\O_A$ satisfying
\eq{
\d \pi_\L^{0,4} \O_A = \d \pi_\L^{3,1} \O_A = 0 \;.
}
We will show that $\O_A$ is given exactly by $S = 1$ as defined in \eqref{sympo}, i.e.,
\eq{\label{yauo2a}
\O_A \equiv e^{ i (\vt + \check{\vt})}  \sqrt{\text{det}(\r)} \O = \zeta^1 \wedge \zeta^2 \wedge \zeta^3 \wedge \zeta^4 \;.
}
In order to show this, we first examine how the exterior derivative acts on $\O$, then consider how this changes under the above rescaling.

First, note that on fiber-invariant forms the exterior derivative acts as $\d = \d r^j  \frac{\p}{\p r^j}$. As a consequence, $\d$ acts with respect to the polarization as $ \d : \O_\bcal^{(a,b)^\dcal}(M_8) \rightarrow \O_\bcal^{(a,b+1)^\dcal}(M_8)$. The result is that
\eq{
\d \pi_\dcal^{(0,4)} \O &= \pi_\dcal^{(1,4)} \d \O \\
\d \pi_\dcal^{(3,1)} \O &= \pi_\dcal^{(4,1)} \d \O  \;.
}
Let us decompose the RHS of
\eq{
\d \O &= W_2 \wedge J + W^*_5 \wedge \O
}
with respect to the real polarization as defined by \eqref{rp1}, \eqref{rp2}. Taking
\eq{
W_2 = \frac{1}{2!} W_{2[jk]l} \z^j \wedge \z^k \wedge \bar{\z}^l
}
and using \eqref{sympj}, \eqref{sympo}, this yields
\eq{\label{eq:proj}
\pi_\dcal^{(1,4)} \d \O &= \frac{1}{4!} \left( 6 W_{2[jkl}h_{m]n}
                        -  i  S  \epsilon_{jklm}W_{5n}^*  + 4i S  \epsilon_{n[jkl}W_{5m]}^*\right)
\delta\t^j \wedge \delta \t^k \wedge\delta \t^l \wedge \delta \t^m \wedge \r^n \\
\pi_\dcal^{(4,1)} \d \O &= - \frac{1}{4!}i \left(6  W_{2[jkl}h_{m]n}
                        + i  S \epsilon_{jklm}W_{5n}^* -  4 i S  \epsilon_{n[jkl}W_{5m]}^* \right)
\r^j \wedge \r^k \wedge\r^l \wedge\r^m \wedge \delta\t^n \;.
}
For the first equation, the RHS vanishes if and only if $W_{2[jkl]} = W_5 = 0$; this follows by noting that the first term is symmetric in $m,n$ whereas the last two terms are not\footnote{Note that $W_{2[jkl]}$ can be considered as $\pi^{3,0}_\dcal W_2$ and has 4 degrees of freedom, compared to the total 20 of $W_2$.}. Similarly, the second equation is trivial if and only if $W_{2[jkl]} = W_5 = 0$. Hence $ \d \pi_\dcal^{(0,4)}  \O =0\iff \d \pi_\dcal^{(3,1)} \O=0$. Constructing an $\O_A$ which satisfies this condition is equivalent to choosing the correct scaling factor $S$, which enters in the above equations not just manifestly, but also via $W_{2,5}$.

We are now ready to show that this scaling factor is $S=1$. Indeed, by construction, \eqref{yauo2a} leads to
\eq{
\pi^{0,4}_\dcal \O_A = \delta \t^1 \wedge \delta \t^2 \wedge \delta \t^3 \wedge \delta \t^4 \;.
}
The constraint \eqref{eq:argh} is exactly $\d \delta \t^j = 0$. Hence, $\O_A$ satisfies
\eq{
\d \pi^{0,4}_\dcal \O_A = 0 \;.
}
But then, due to \eqref{eq:proj}, it follows that also
\eq{
\d \pi^{3,1}_\dcal \O_A = 0
}
and hence the theorem is proven. }

Some thoughts on the constraint \eqref{eq:argh}. The reason this issue does not arise in \cite{lty}, is because there, $\O_A$ is constructed by mirror symmetry and automatically satisfies $A^j_{\phantom{j}k} = 0$. In a sense, the trivial connection is related to the fact that the NSNS-form $\check{H}$ on the 2B side ({\it not} the 2A side!) is trivialized in their case. This can be seen from the definition of T-duality, which will be examined in the next section; see the discussion below \eqref{uniconnection}. The constraint can be rewritten in terms of torsion classes, specifically, $W_2$ and perhaps $W_1$ as well, by means of a long and tedious computation that we have not undertaken.

\section{Mirror symmetry}\label{ms}
So far, we have simply discussed the LTY 2A and 2B systems, and compared them with certain supersymmetric solutions. The LTY 2B system captured some, but not all the constraints of the complex supersymmetric system, whereas on the other hand, the LTY 2A system captures some of the symplectic supersymmetric system, but additionally also imposes some constraints which are not present in supersymmetry, in particular, the necessity of a special real polarization and \eqref{eq:argh}. The LTY 2A and 2B systems are mirror systems in a sense that we will make precise. Similarly, the complex and symplectic supersymmetric systems are mirror symmetric in another sense that is to be made precise. Thus, a natural question would be: Why is the LTY 2A system not simply a subset of the symplectic supersymmetry constraints? The answer has to do with the fact that the LTY mirror map is constructed in such a way that complex geometry is mapped to symplectic geometry. However, this is not quite the natural framework of supersymmetry. Instead, the correct framework is that of generalized complex geometry; in particular, the cases we have been considering are those where the the generalized complex structure is induced either by a complex or symplectic structure. There is a map which preserves the generalized complex structures instead, namely T-duality, which has been discussed in this context in \cite{bem}, \cite{bhm}, \cite{cg}.

In this section, we will discuss the particularities of mirror symmetry a la LTY and discuss how the 2A and 2B system are miror symmetric. We then discuss T-duality, compare it with the LTY mirror map, and demonstrate how T-duality relates the complex and symplectic supersymmetric solutions to one another, whereas the LTY mirror map does not.

Following \cite{lty} we denote the relation between the LTY systems as `mirror symmetry', whereas we use `T-duality' to denote the relation between the supersymmetric systems.

\subsection{LTY Mirror symmetry}
We begin by discussing in what sense LTY 2A and LTY 2B are mirror symmetric to each other, as discussed in \cite{lty}.
We consider the setup as discussed in section \ref{tb}, with $(M_A, J_A)$ a Lagrangian torus fibration over $\bcal$ and $(M_B, \O_B)$ the complex dual torus bundle with complex structure induced by $\O_B$.

\dfn{$M_B$ is equipped with complex and real polarities, induced by the complex structure and fibration respectively. We define the {\it polarity switch operator}
\eq{
P : \O_\bcal^\bullet (M_B)\rightarrow \O_\bcal^\bullet (M_B)
}
as follows. For any fiber-invariant form
\eq{
\a = \a_{j_1...j_p k_1...k_q} (r) \d z^{j_1} \wedge ... \wedge \d z^{j_p} \wedge \d \bar{z}^{k_1} ... \wedge \d \bar{z}^{k_q}
}
we set
\eq{\label{pso}
P(\a) \equiv \a_{j_1...j_p k_1...k_q} (r) \d \check{\t}^{j_1} \wedge ... \wedge \d \check{\t}^{j_p} \wedge \d r^{k_1} ... \wedge \d r^{k_q}\;,
}
with $\check{\t}^j$ the fiber coordinates of $M_B$.
}

Note that, as can be shown by making use of \eqref{cplxcoords}, the polarity switch operator is in fact an isomorphism between differential complexes:
\eq{
(\O^\bullet_\bcal(M_B), \p, \bar{\p}) \simeq (\O^\bullet_\bcal(M_B),  \frac{i}{2} \d^c,   \frac{i}{2}  \d )  \;.
}
This will play an important role in section  \ref{tduality}. By making use of the polarity switch operator, we can define the Fourier-Mukai transformation which maps forms on $M_A$ to $M_B$.

\dfn{
Let
\eq{
\tilde{f} \equiv  \d \check{\t}^j \wedge \d \t^j
}
be the `curvature of the universal connection' (up to prefactor) on the correspondence space $M_A \times M_B$.
Then
the {\it Fourier-Mukai transform} is defined as
\eq{\label{fm}
FT &: \O^\bullet (M_B) \rightarrow \O^\bullet (M_A) \\
FT (\a) &= \check{\pi}_* \left( P(\a) \wedge \exp{\tilde{f} }\right).
}
Here, $\check{\pi}_*$ is defined as integration over the fibers of $M_B$, which are locally parametrized by the coordinates $\check{\t}^j$.
}
Using this definition, the Fourier-Mukai transform is invertible. In fact, it is an isomorphism of the differential complexes of $M_A$ and $M_B$:
\eq{\label{fmiso}
(\O^\bullet_\bcal(M_B), \p, \bar{\p} ) \simeq (\O^\bullet_\bcal(M_A), (-1)^n \frac{i}{2} \d^\L,  (-1)^n \frac{i}{2}  \d ) \;.
}
This isomorphism is why the polarity switch operator is used in order to define the Fourier-Mukai transform; it translates the natural derivatives of complex and symplectic geometry into each other. Making use of this definition of the Fourier-Mukai transform, the LTY 2A and 2B systems are related as follows.

\thm{\label{yausthm}
Let $J_B$ be a fiber-invariant real positive-definite (1,1)-form on $M_B$ and $\O_A = FT (\exp(2 J_B))$. Then we have the following:
\begin{enumerate}
\item $(M_B, J_B, \O_B)$ is an $SU(n)$-structure manifold $\iff$ $(M_A, J_A, \O_A)$ is an $SU(n)$-structure manifold. If so, $K_B =  K_A^{-1}$.
\item Moreover, $(M_B,J_B, \O_B)$ satisfies \eqref{y2b} if and only if $(M_A, J_A, \O_A)$ satisfies \eqref{y2a}.
\item In case 2. holds, $\exists k \in \cbb$ such that $\r_A = k~ FT (\r_B)$.
\end{enumerate}
}
In this sense, the LTY 2A and 2B systems are mirror symmetric.

\subsection{Mirror symmetry for the supersymmetric systems}
We have now seen how the LTY 2A and 2B systems are mirror symmetric to each other.
Next, let us investigate in what sense the complex and supersymmetric systems are mirror symmetric to each other.

Both the complex and symplectic supersymmetric systems are described by \eqref{susy}. The difference is that for complex supersymmetry, one has that
\eq{\label{cplxpoly}
\check{\Psi}_{2} &= - e^{- i \check{\vt}} e^{-i\check{J}} \\
\check{\Psi}_{1} &= - e^{i \check{\vt}} \check{\O}
}
whereas in the symplectic case, one has
\eq{\label{symppoly}
\Psi_{1} &= - e^{i \vt} \O\\
\Psi_{2} &= - e^{- i \vt} J
}
with checks added for clarity on the complex side. Thus, `mirror symmetry' between these systems should (very roughly) be given by $\check{\Psi}_{j} \leftrightarrow \Psi_{j}$ i.e., interchanging the polyforms, which determine the profiles for the fluxes. Note that no restriction whatsoever is placed on the geometry at this point, other than admitting an $SU(4)$-structure in the first place.

To be more precise, we are looking for some non-trivial map $T$ acting on polyforms, such that the constraints \eqref{susy} are mapped to themselves. Generically, the outcome need not be the supersymmetry constraints manifestly, but we will limit ourselves to looking for a map $T$ which satisfies
\eq{
T \left[ \d_{\check{H}} \left( e^{2\check{A} - \check{\phi}} \text{Re} \check{\Psi}_1 \right)\right] &= \d_H \left( e^{2A - \phi} \text{Re} \Psi_1 \right) \\
T\left[ \d_{\check{H}} \left( e^{2 \check{A} - \check{\phi}} \check{\Psi}_2 \right)\right] &= \d_H \left( e^{2 A - \phi} \Psi_2 \right) \\
T\left[ \d_{\check{H}}^{\check{\jcal}_2} \left( e^{- \check{\phi}} \text{Im} \check{\Psi}_1 \right)\right] &= \d_H^{\jcal_2} \left( e^{- \phi} \text{Im} \Psi_1 \right)\;.
}
This map will be given by the T-duality map, which will be defined by \eqref{t}, provided we take $\check{A} = A$ and assume that the complex system lives on $M_B$, whereas the symplectic system lives on $M_A$, with the appropriate torus bundle structures on $M_A, M_B$. Provided this is satisfied, the T-duality transformation of the fluxes follows, analogously to the $d=4$ case in \cite{gmpw}.

\subsection{T-duality}\label{tduality}
In this section, we discuss the T-duality map that provides the correct notion of mirror symmetry with respect to the supersymmetric systems. We mostly follow along the lines of \cite{cg}.

We start by considering two manifolds $M$, $\check{M}$ which are two torus fibrations over $\bcal$. Generically, T-duality can be defined without the need to specify symplectic or complex structures on $M$, $\check{M}$, and their fibers need not be dual.
\dfn{
Let $M, \check{M}$ be equipped with fiber-invariant three-forms $H, \check{H}$. Suppose
\eq{\label{uniconnection}
\d f = p^* H   - \check{p}^* \check{H} \;,
}
with $p : M \times \check{M} \rightarrow M$ the natural projection operator, and $f\in \O^2_\bcal (M \times \check{M})$ non-degenerate when acting on the tangent space to the fibers\footnote{In particular, $\tilde{f}$ is such a map, but we will have to generalize $\tilde{f}$ to incorporate the non-triviality of $\check{H}$.}. Then $(M, H)$ will be called {\it T-dual} to $(\check{M}, \check{H})$.
}
This definition restricts $H, \check{H}$ as follows. Given fiber coordinates $\{\t^j\}$, $\{\check{\t}^j\}$, we can locally set
\eq{\label{uniconnection2}
f &=  (\d \t^j + A^j_{\phantom{j}l}(r) \d r^l ) \wedge (\d \check{\t}^k + \check{A}^k_{\phantom{k}m}(r) \d r^k)\\
&\equiv  \delta \t^j \wedge \delta \check{\t}^j
\;.
}
Thus, we see that
\eq{\label{Hdef}
\check{H} &=  \check{\hcal}  +  \hcal^{(\bcal)} \\
H &=   \hcal +  \hcal^{(\bcal)}\;,
}
with $\hcal \in \O^2(\bcal; \mathfrak{t}^*)$, $\check{\hcal}\in \O^2(\bcal; \check{\mathfrak{t}}^*)$ locally given by
\eq{\label{Hdef2}
\check{\hcal} &= - \d \delta \t^j \wedge \delta \check{\t}^j = \check{\hcal}_{jkl}\delta \check{\t}^j \wedge \d r^k \wedge \d r^l \\
\hcal &= - \delta \t^j \wedge \d \delta \check{\t}^j  = \hcal_{jkl}\delta \t^j \wedge \d r^k \wedge \d r^l \;,
}
and $\hcal^{(\bcal)}$ a basic form, i.e., a form on $\bcal$.
In the case of dual fibers, the connections satisfy $\delta \t^j \in \O^1(M; \mathfrak{t})$, $\delta \check{\t}^j \in \O^1(\check{M}; \mathfrak{t}^*)$, hence the sum over $j$ should be considered as the natural pairing of $\mathfrak{t}^*$ with $\mathfrak{t}$.

Given two such T-dual spaces, we can define the following map
\dfn{
The T-duality map is defined as
\eq{\label{t}
T : \O^\bullet_\bcal (\check{M}) \rightarrow \O^\bullet_\bcal (M) \\
T(\a) \equiv \check{\pi}_\star \left( \a \wedge \exp{f }\right)\;.
}
}
The definition of $f$ and the resulting constraints on $H$, $\check{H}$ are constructed exactly such that
\eq{
T (\d_{\check{H}} \a ) = \d_H T(\a) \qquad \forall \a \in \O_\bcal^\bullet(\check{M}) \;.
}
Comparing the definitions for the T-duality map and the Fourier-Mukai transform \eqref{fm} for the case $H=\check{H}=0$, we see that the difference is exactly the polarization switch operator, that is\footnote{Up to a conventional sign difference between $f$, $\tilde{f}$.},
\eq{
FT = T \circ P \;.
}
The Fourier-Mukai map gives the isomorphism
\eq{\label{fmiso2}
\left(\O^\bullet_\bcal(M_B), \p, \bar{\p} \right) \simeq \left(\O^\bullet_\bcal(M_A),  (-1)^n \frac{i}{2} \d,  (-1)^n\frac{i}{2}  \d^\L \right) \;,
}
as noted in \eqref{fmiso}. This is not the case for this T-duality map. Instead, the T-duality map induces a map between Courant algebroids, such that, given generalized complex structures on $\check{M}$, these are mapped to generalized complex structures on $M$. As a result, the T-duality map
gives an isomorphism between the complexes
\eq{\label{tiso}
(\O^\bullet_\bcal(\check{M}), \p^{\check{\jcal}}, \bar{\p}^{\check{\jcal}} ) \simeq (\O^\bullet_\bcal(M), \p^\jcal,  \bar{\p}^\jcal ) \;,
}
as shown by Thm (4.1, 4.2) of \cite{cg}.

Let us investigate this in more detail. Consider now the setup of \ref{yausthm} with $M = M_A$ a Lagrangian torus fibration and $ \check{M} = M_B$ the complex dual torus fibration. In this case, both $M$ and $\check{M}$ come with an integrable generalized almost complex structure, and a non-integrable generalized almost complex structure.
Since the generalized complex structure of $M$ is associated to the symplectic structure ($\jcal = \jcal_J$), and the generalized complex structure of $\check{M}$ is associated to the complex structure ($\check{\jcal} = \check{\jcal}_I$), it follows that T-duality gives us the isomorphism
\eq{
(\O^\bullet_B(\check{M}), \p, \bar{\p} ) \simeq (\O^\bullet_B(M), \p^{\jcal_J},  \p^{\jcal_J} ) \;.
}
We see that on the lefthand side, the differentials reduce to the ordinary Dolbeault operator and its conjugate, i.e., $\p = \p^{\jcal_I}$. However, the differentials on the righthand side are {\it not} the $\d$ and $\d^\L$. Instead, the generalized Dolbeault operator associated to the symplectic generalized complex structure is a more complicated thing, as described in \cite{cavalcanti}. Thus, one has the following commutative diagram:
\eq{
\xymatrix{
\left(\O^\bullet_\bcal(M_B), \p, \bar{\p} \right)  \ar[d]^P \ar[rd]^{FT}  \ar[r]^T &  \left(\O^\bullet_\bcal(M_A),  \p^{\jcal_J},  \bar{\p}^{\jcal_J} \right) \ar[d]^Q \\
\left(\O^\bullet_\bcal(M_B), \frac{i}{2} \d^c, \frac{i}{2} \d \right) \ar[r]^T &  \left(\O^\bullet_\bcal(M_A), \frac{i}{2} \d^\L, \frac{i}{2} \d \right)
}
}
The isomorphism $Q$ was found in \cite{cavalcanti} but does not play an important role for us.

Let us consider this from a different angle by examining how the polyforms given in \eqref{cgacs} behave under Fourier-Mukai and T-duality maps.
Up to a rescaling of the polarization switch operator, we have that the polyform associated to the non-integrable generalized almost complex structure satisfies
\eq{
FT( \check{\Psi}_1) = \Psi_1 \;.
}
What about the other polyform? Up to a scalar factor, $\check{\Psi}_2$ is given by
\eq{
\check{\Psi}_2 \sim  \O_B \sim \d z^1 \wedge \d z^2 \wedge \d z^3 \wedge \d z^4 \;,
}
with holomorphic one-forms given by \eqref{cplxcoords}.
Thus, we can explicitly compute
\eq{
FT (\check{\Psi}_2) \sim 1   \;.
}
This pure spinor is associated to the subbundle $L = T \otimes \cbb \subset (T \oplus T^*) \otimes \cbb $, which is maximal and isotropic, but does not satisfy $L \cap \bar{L} = \{0\}$ and hence, is not an (almost) Dirac bundle. In other words, the image of $\check{\Psi}_2$ is not associated to a generalized almost complex structure.

Now let us now see how the polyforms behave under T-duality. We restrict to the relevant case, where the polyforms are determined by the Killing spinors and satisfy the supersymmetry equations \eqref{susy}. The Killing spinors yield \eqref{cplxpoly} and the (complex) fibration structure of $M_B$
yields \eqref{cplxo}, \eqref{cplxj}. Noting that the scale factor $\check{S}$ of $\O_B$ is determined by  $W_5$ given in \eqref{pt}, that $A = \check{A}$, and that $\p \vt =0$, we find that
\eq{
\check{\Psi}_{1} &=  - e^{ -i \vt} e^{\check{\phi} - 2A} \d z^1 \wedge \d z^2 \wedge \d z^3 \wedge \d z^4 \\
\check{\Psi}_{2} &= - e^{- i \check{\vt}} \exp\left({\frac12 \check{h}_{jk} \d z^j \wedge \d \bar{z}^k   } \right)\;.
}
Furthermore, triviality of the conformal factor \eqref{cplxconf} gives us a link between the fields and the geometry:
\eq{\label{link}
\sqrt{\text{det}\check{h}} = e^{\check{\phi} - 2A} \;.
}
Triviality of $H$ of the symplectic supersymmetry solution \eqref{eq:sympsusy} determines that we should consider\footnote{Actually, we could consider any $\delta \check{\t}^j$ satisfying $\d \delta \check{\t}^j = 0$, but we will neglect this for convenience.}
\eq{
f = \d \t^j \wedge \d\check{\t}^j \;.
}
Making use of \eqref{sympj}, \eqref{sympo}, \eqref{symppoly}, we then find that
\eq{
T(\check{\Psi}_1) &= \sqrt{\text{det}\check{h}} \Psi_1 \\
T(\check{\Psi}_2) &= \sqrt{\text{det}\check{h}} \Psi_2\;,
}
where we have identified
\eq{
\delta_{jl} \r^l_{\phantom{l}k} = \check{h}_{jk} \;.
}
Inserting \eqref{link} and noting that the symplectic supersymmetry systems \eqref{eq:sympsusy} requires $A = \frac12 \phi$, we thus have that
\eq{\label{tpoly}
T(\check{\Psi}_1) &= e^{\check{\phi} - \phi} \Psi_1 \\
T(\check{\Psi}_2) &= e^{\check{\phi} - \phi} \Psi_2\;,
}
which is exactly the result that maps the complex supersymmetric system into the symplectic supersymmetric system. Note that the non-constant prefactors
came about through an intricate interplay between the fibration structures and supersymmetry; up to constant scalars, we did not put the `correct' result in by hand. Furthermore, note that the generalized almost complex structure associated to $\Psi_1$ gets mapped to another generalized almost complex structure, despite non-integrability.

Note also that the restrictions on $H$, $\check{H}$ coming from the fibration structure, namely \eqref{Hdef}, \eqref{Hdef2} are compatible with the demands on $H$, $\check{H}$ from supersymmetry for both the symplectic and complex case. For the symplectic supersymmetric system \eqref{symp}, $H=0$, so this statement is evident.
For the complex supersymmetric system, $\check{H} = h^{(2,1)} + h^{(1,2)}$. By making use of the holomorphic one-forms \eqref{cplxcoords} on $M_B$, it follows from \eqref{Hdef2} that $\check{\hcal} \in \O^{(2,1)}(M_B) \oplus \O^{(2,1)}(M_B)$, whereas $\hcal^{(\bcal)}$ needs to be trivial due to vanishing of $H$. Primitivity of $\check{H}$ then comes down to
\eq{
\left(\check{h}^{-1}\right)^{jk} \p_{[l} \check{h}_{j]k} = 0\;,
}
which can be considered a constraint on the fibration structure (noting that $\check{h}$ is related to $\r$) from supersymmetry.

To summarize, we have found the following. By construction of the T-duality map, $\d_H$ is preserved, and by assumption the warp factor is equivalent on both sides. Furthermore, we have seen that the polyforms $\Psi_j$ associated to the generalized almost complex structures are mapped into each other with the correct prefactor, i.e., \eqref{tpoly}. Thus, looking at the supersymmetry equations \eqref{susy}, we conclude that
\eq{
T\left( \d_{\check{H}} \left( e^{2\check{A} - \check{\phi}} \text{Re} \check{\Psi}_1 \right)\right) &=
\d_H \left( e^{2A - \phi} \text{Re} \Psi_1 \right)  \\
T\left( \d_{\check{H}} \left( e^{2 \check{A} - \check{\phi}} \check{\Psi}_2 \right)\right) &=
\d_H \left( e^{2 A - \phi} \Psi_2 \right) \\
T\left(  \d_{\check{H}}^{\check{\jcal}_2} \left( e^{- \check{\phi}} \text{Im} \check{\Psi}_1 \right)\right) &=
\d_H^{\jcal_2} \left( e^{- \phi} \text{Im} \Psi_1 \right)
}
exactly as desired. In other words, this definition of T-duality is the one that correctly maps complex supersymmetric solutions onto symplectic supersymmetric solutions.

In fact, the relation is deeper. Not only do supersymmetric solutions get mapped to supersymmetric solutions, but backgrounds are mapped to backgrounds as well. Let us recall that the supergravity equations of motion are satisfied by a supersymmetric solution if this solution satisfies the integrability conditions: the Bianchi identities are satisfied and the external $B$-field equation is satisfied. For both the complex and symplectic supersymmetric solutions, these are studied more in-depth in appendix \ref{integ}. In both cases, the external $B$-field equation is trivial. Let us show that satisfaction of the Bianchi identities is preserved under the T-duality map.

The Bianchi identities in our conventions are given by \eqref{bianchi}. Let us first consider the NSNS Bianchi identity. Making use of  \eqref{uniconnection}, \eqref{uniconnection2} and \eqref{Hdef}, it is straightforward to verify that $\d H = 0$ if and only if $\d \check{H} = 0$. In fact, since the symplectic supersymmetry solution requires $H=0$, we immediately see that both NSNS Bianchi identities are satisfied automatically in our setup.
Next, let us consider the RR Bianchi. Using the decomposition into electric and magnetic components as given by \eqref{fluxdecomp}, the Bianchi identity
for the RR fluxes reduces to
\eq{
\d_H e^{2A} \star_8 \s F = \d_H F = 0 \;.
}
The NSNS Bianchi identity is satisfied and hence $\d_H^2 = 0$. Then the supersymmetry equations \eqref{susy} imply that $\d_H \star_8 \s F = 0$ holds automatically, whereas $\d_H F = 0$ can be translated to the purely geometrical constraint
\eq{
\d_H \d_H^{\jcal_2} \left( e^{- \phi} \Im \Psi_1\right) = 0 \;.
}
Since the T-duality map preserves the generalized almost complex structures, the NSNS flux, and the polyform $e^{-\check{\phi}} \check{\Psi}_1$, we find that
\eq{
\d_{\check{H}} \d_{\check{H}}^{\check{\jcal}_2} &\left( e^{- \check{\phi}} \Im \check{\Psi}_1\right)  = 0  \\
&\iff \\
\d_H \d_H^{\jcal_2} &\left( e^{- \phi} \Im \Psi_1\right)
= T \Big[\d_{\check{H}} \d_{\check{H}}^{\check{\jcal}_2} \left( e^{- \check{\phi}} \Im \check{\Psi}_1\right) \Big] =  0 \;.
}
This shows that indeed, the integrability conditions for one supersymmetric solution are satisfied if and only if they are satisfied for the mirror.

\section{F-theory backgrounds}
\label{ftheo}
The solution of section \ref{cplxsusy} is specified by a relatively large number of parameters, and is subject to tadpole conditions.  One way of fixing a number of parameters is to enforce integrability in the absence of sources. As discussed in appendix \ref{integ},  such solutions involve conformally Calabi-Yau fourfold manifolds with NS three-form and RR five-form internal fluxes. In this section, we shall briefly consider solutions that admit F-theory lifts. As we shall see they are also related to a special choice of parameters and  involve  conformal Calabi-Yau fourfolds.  The discussion of tadpoles is much more convenient in M/F-theory language.

 F-theory on Calabi-Yau fivefolds is not much studied (see \cite{hls, snw} however).  While we do not directly analyse supersymmetry conditions of M-theory, we show that it is possible to restrict the complex supersymmetric system in such a way that the lift is possible in a fashion very similar to the more familiar case of IIB backgrounds with primitive self-dual three-form flux on confomally Calabi-Yau threefolds.

The familiar case of F-theory on an elliptically fibered Calabi-Yau fourfold $Y_4 \rightarrow \bcal_3$ is related to type IIB on the base space $\bcal_3$ of the elliptic fibration in the presence of D7-branes. The base space $\bcal_3$ is a K\"{a}hler manifold which does not admit an $SU(3)$-structure and in general is not even a spin manifold. However, considering  its  double cover, branched along the divisor wrapped by an O7-plane, yields a Calabi-Yau threefold $Z_3$ and yields a good weakly-coupled description of IIB compactifications \cite{gkp, gp}. Such backgrounds satisfy the following conditions: The metric is given by
\eq{
ds^2(M_{10}) = e^{2A} ds^2(\rbb^{1,3}) + e^{-2A} ds^2(Z_3) \;,
}
where $ds^2(Z_3)$ is the Ricci-flat Calabi-Yau metric. The axio-dilaton $\tau = C_0 + i e^{-\phi}$ is holomorphic and the three-form flux $G_3 = \fcal_3 - \tau H$ is primitive (2,1) and imaginary selfdual. O7 panes are needed for allowing compact $Z_3$, and D3/O3 sources are allowed. In M/F-theory language, the primitivity of $G_3$ flux translates into the familiar primitivity of the internal component of the four-form-flux $G_4$ \cite{bb, gvw}, and the tadpole condition in the absence of M2/D3-branes becomes
\eq{
\frac{1}{8 \pi^2} \int_{Y_4} F \wedge F = \frac{\chi(Y_4)}{24}\;,
}
where $F$ is the internal component of $G_4$ and $\chi$ the Euler number.

The complex supersymmetric system \eqref{pt} has a special case that is a direct analogue of this solution. Considering
\eq{
\vt &= - \frac{\pi}{2} \\
e^\phi &= g_s e^{-2A} \\
f^{(2,1)}_5 &= 0 \;
}
yields the following intrinsic torsions:
\eq{
W_1 &= W_2 = W_3 = 0 \\
W_4 &= \frac{1}{2} W_5 = \p \phi \;.
}
Hence the  ten-dimensional metric is given by
\eq{
ds^2(M_{10}) = e^{-\phi} ds^2(\rbb^{1,1}) + e^{\phi} ds^2(Z_4)
}
for some Calabi-Yau fourfold $Z_4$, and the fluxes reduce to
\eq{
\fcal_1 &= - \d^c e^{- \phi} \\
\fcal_3 &=  f_3^{(2,1)} + \text{c.c} \\
\fcal_5 &= \frac12 \d^c e^{- \phi} \wedge J^2 - \vol_2 \wedge \d e^{- \phi} \wedge e^{- \phi} J \\
H &= - i e^\phi f^{(2,1)}_3 + \text{c.c.}
}
Therefore, it follows that
\eq{
\bar{\p} \left(C_0 + i e^{-\phi} \right) &= 0 \\
G_3 &= i C_0 e^\phi \left(f_3^{(2,1)} - f_3^{(1,2)}\right)\;.
}
In particular, we see that $\tau$ is holomorphic and that $G_3$ is primitive (2,1) and not invariant under $SL(2,\zbb)$ transformations, exactly as desired. $f_3^{(2,1)}$ is not fixed by supersymmetry, but subject to tadpole cancellation constraints.

Once more thinking of $Z_4$ as the double cover of a K\"ahler manifold $\bcal_4$, branched along
the divisor wrapped by an O7-plane, we can lift this solution to a compactification of M/F-theory on (conformally) elliptically fibered fivefold $Y_5 \rightarrow \bcal_4$ with the internal part of the  four-from flux that is $(2,2)$ and primitive:
\eq{ F\wedge J_{Y} \wedge J_{Y} = 0 \,.}
As follows from the M-theory flux equation of motion, the eight-form given by
\eq{
\eta &= \frac{1}{8 \pi^2}  F \wedge F  - X_8
}
is trivial in cohomology in absence of sources and should integrate to zero on any complex four-cycle in $Y_5$. Here, $X_8$ is a higher order correction, given in terms of the the Pontryagin classes as \cite{dlm}
\eq{
X_8 = \frac{1}{48} \left(p_2 - \frac{1}{4} p_1^2 \right)\;.
}
Note that the external Einstein equation will follow once the  tadpole conditions are satisfied, since in absence of sources it reads:
\eq{
\frac{1}{8 \pi^2} \int_{Y} F \wedge F \wedge J_Y  = \int_{Y} X_8 \wedge J_Y \;.
}

\noindent
It will be of some interest to study the existence of compact string backgrounds corresponding to generic complex and symplectic systems, given by  \eqref{pt} and \eqref{eq:sympsusy} respectively.

\section*{Acknowledgments}

We would like to thank Raffaele Savelli, Li-Sheng Tseng and Dimitrios Tsimpis for helpful discussions. RM would like to thank KIAS for hospitality. This work was supported in part by the Agence Nationale de la Recherche under the grant 12-BS05-003-01.

\appendix

\section{Integrability} \label{integ}
In order to solve the equations of motions, a supersymmetric solution needs to satisfy the integrability conditions\footnote{Not to be confused with integrability of any geometrical structure.}. In this section we give a number of simple examples that do so. The example for the complex supersymmetric solution was derived in \cite{stenzel}, the ones for the symplectic solution are new.

\subsection{Backgrounds from the complex supersymmetric solution} \label{ccy}
Consider the complex supersymmetry solution \eqref{pt} and set
\eq{
\vt =& \pi\\
e^{\phi} =& g_s e^{-2A}\\
f_3^{(2,1)} =& 0 \;.
}
As a consequence, the torsion classes are given by
\eq{\label{eq:iibtors}
W_1 = W_2 = W_3 &= 0 \\
W_4 = \frac12 W_5 &= - 2 \p A \;,
}
which gives us a conformal Calabi-Yau structure, with the conformal metric $g_8$ related to a Calabi-Yau metric $g_{CY}$ by
\al{\label{eq:ccymetric}
g_8 = e^{-2 A } g_{CY} ;.
}
The NSNS three-form is given by
\al{\label{eq:iibnsfluxes}
H = h^{(2,1)} + h^{(1,2)}\;,
}
in particular, $H$ is internal and primitive.
The non-vanishing RR fluxes are given by
\eq{\spl{\label{eq:iibrrfluxes}
{g_s}\fcal_3&= \mathrm{vol}_2\wedge\d e^{4A}\\
{g_s}\fcal_5&= e^{4A}\mathrm{vol}_2\wedge H- e^{2 A}\star_8 H\\
g_s \fcal_7 &= \star_{10} \s \fcal_3\;.
}}
In this case, the vanishing of the external $B$-field eom is automatic. The Bianchi identities lead to a constraint on the warp factor.
The non-trivial Bianchi identies are given by
\eq{
\d H  &= 0 \\
\d \fcal_5 + H \wedge \fcal_3 &= 0\\
\d \fcal_7 + H \wedge \fcal_5 &= 0 \;.
}
The second line implies
\al{ \label{eq:iibbianchis}
\d H = \d e^{2 A}\star_8 H = 0
}
while the third implies
\al{
\d \star_8 \d e^{2 A} +\frac{1}{2} H \wedge e^{2 A} \star_8 H = 0 \;.
}
In terms of the Calabi-Yau metric, these can be rewritten as respectively
\eq{
\d H = \d \star_{CY} H = 0
}
and
\al{\label{eq:iibwarp}
- \d \star_{CY} \d e^{-4 A} + H \wedge \star_{CY} H = 0 \;.
}
The background is parametrized by $h^{(2,1)}$ and $A$. One can consider an even simpler yet still non-trivial subcase by taking $h^{(2,1)}=0$.

\subsection{Backgrounds from the symplectic supersymmetric solution}
In the case of the sympletic supersymmetric solution, the integrability conditions are as follows. Firstly, we again find that the vanishing of the external part of the $B$-field equation of motion is automatic. Hence the only integrability condition is the Bianchi identity for the fluxes. Taking into account the vanishing of $H, F_{1,5,7}$, these reduce to
\eq{
\d F_3 = \d e^{2A} \star_8 F_3 = 0 \;,
}
with
\eq{
F_3 = \tilde{f}_3^{(0,1)} \lrcorner \O + f_3^{(2,1)} + \text{c.c.}
}
We will give two simple solutions to this constraint.

The first is obtained by setting
\eq{
\tilde{f}_3^{(1,0)} = \d f^{(2,1)} = \d A = 0 \;.
}
As a result the dilaton and $W_5$ are trivial. The geometry of the background is nearly Calabi-Yau, i.e., the manifold is symplectic but not complex, with $W_1=W_3=W_4=W_5 = 0$. The background is fully parametrized by a single closed primitive (2,1)-form flux $f_3^{(2,1)}$.

The second solution is given by
\eq{
f_3^{(2,1)} = 0 \;.
}
In this case, the Bianchi identity can be rewritten as
\eq{
\d \left( \tilde{f}_3^{(0,1)} \lrcorner \O \right) &= 0\\
(2 \d A \wedge \tilde{f}_3^{(0,1)} + \d \tilde{f}_3^{(0,1)}) \wedge \O &= 0
}
Noting that $\d \O \sim \bar{\p} A \wedge \O$, $\tilde{f}_3^{(0,1)} \sim \bar{\p} \log A$, the second equation follows immediately. The first one follows due to the fact that
\eq{
0&= \d \left( \bar{\p} A \lrcorner \O \right)  \\
&\iff \\
0 &=  \d \left( \bar{\p} A \lrcorner \O \wedge J \right)
\sim \d \left( \bar{\p} A \wedge \O \right) \sim \d^2 \O\;.
}
The resulting background is K\"{a}hler and fully parametrized by the single real scalar function $A$.

\subsection{On conformal Calabi-Yau backgrounds}
In this section, we discuss a curious feature exhibited by the conformal Calabi-Yau background discussed in \ref{ccy}. The conformal Calabi-Yau background was obtained by imposing certain conditions on the complex supersymmetric system of section \ref{cplxsusy}, which are such that the integrability conditions are satisfied. In particular, the Bianchi identities are satisfied and hence
\eq{
\d_H \fcal = 0 \implies \d_H F \equiv \r_B = 0\;.
}
By theorem \ref{yausthm}, it follows that the mirror of such a conformal Calabi-Yau background satisfies $\r_A=0$. But in the case that $\r_A = 0$, one might hope to find an $SU(4)$-structure $(J_A, \O_A, K_A)$ satisfying the LTY 2A system \eqref{y2a} such that
\eq{\label{ccymiracle}
\d \O_A = \d K_A = 0 \;,
}
in which case the entire story of section \ref{2asymp} is simplified greatly; one no longer needs the fibration structure and can simply compare the LTY 2A system to the supersymmetric system directly, just like the case for the complex supersymmetric system/LTY 2B comparison.
Strangely enough, it turns out that such a supersymmetric system is exactly the conformal Calabi-Yau itself! More concretely, the conformal Calabi-Yau, which is a particular complex supersymmetric system, yields both a solution to the LTY 2B {\it and} to the LTY 2A system.
That the conformal Calabi-Yau leads to a set $(J_B, \O_B, K_B, \r_B)$ satisfying $\eqref{y2b}$ is clear, as this is just the story of section \ref{2bcplx}. Given the conformal Calabi-Yau solution, a set  $(J_A, \O_A, K_A, \r_A)$ satisfying \eqref{y2a} is given by
\eq{\label{ccy2a}
J_A &= e^{\frac23 (\phi - A)} \\
\O_A &= e^{- \frac13 \left(5 \phi - 2 A\right)+ i \vt } \O\\
K_A &= e^{- \frac23 \left( \phi + 2A\right)}\\
\r_A &= 0\;.
}
These were constructed as follows. First, one notes that for the conformal Calabi-Yau, $W_1=W_3 = 0$, and hence, in order to construct the symplectic form $J_A$, all one needs to do is rescale $J$ such that $W_4$ vanishes. But this is exactly the balanced condition that was needed when constructing $J_B$, so we just take $J_A = J_B$. Next, recalling that $K_A = K_B^{-1}$ was a result of the Fourier-Mukai map, we take this as definition for the conformal factor $K_A$ and then see what the resulting rescaling of $\O$ is which leads to the definition of $\O_A$. Plugging in the constraint
\eq{
\d (\phi + 2A ) = 0
}
which determines the conformal Calabi-Yau solution, it then follows that \eqref{ccymiracle} is satisfied. As a consequence, \eqref{y2a} is satisfied.

There are two straightforward and probably rather useless generalizations of this construction. Firstly, note that similar to the situation for the complex/2B comparison,  \eqref{ccy2a} can be generalized somewhat by including a number of constant factors. Secondly, note that we have identified $\phi, A$ with $\check{\phi}, \check{A}$; by allowing these to be independent, one may construct a slightly more convoluted $(J_A, \O_A, K_A)$ satisfying \eqref{y2a}.

\section{$SU(4)$-structures}\label{su4decomp}
In this section we discuss the decomposition of forms with respect to $SU(4)$-structures, which is essential to acquire the supersymmetric solutions discussed in section \ref{supersymmetry}.

Given an $SU(4)$-structure, any form can be decomposed into irreps of $SU(4)$. From a more geometric point of view, this is equivalent to Lefschetz and Hodge decomposition of $k$-forms into primitive $(p,q)$-forms. Concretely, we have that one-, two-, three-, selfdual four-, and anti-selfdual four-forms can be globally decomposed as
\eq{\label{eq:su4decomp}
F_m &= f^{(1,0)}_m + \mathrm{c.c} \\
F_{mn} &=f^{(1,1)}_{2|mn}+f_2J_{mn}+\left(f^{(2,0)}_{2|mn}+\mathrm{c.c.}\right) \\
F_{mnp} &=f^{(2,1)}_{3|mnp}+3f^{(1,0)}_{3|[m}J_{np]}+ \tilde{f}^{(1,0)}_{3|s}\Omega^{s*}{}_{mnp} +\mathrm{c.c.}\\
F^+_{mnpq} &=f^{(2,2)}_{4|mnpq}+6f_{4}J_{[mn}J_{pq]} +\left(6f^{(2,0)}_{4|[mn}J_{pq]}+\tilde{f}_4\Omega_{mnps}+\mathrm{c.c.}\right)\\
F^-_{mnpq}&=6f^{(1,1)}_{4|[mn}J_{pq]}+\left(f^{(3,1)}_{4|mnpq}+\mathrm{c.c.}\right)\;.
}
All forms $f^{(p,q)}$ in these expresions are primitive, that is to say, they satisfy
\eq{
f^{(p,q)}_{m_1...m_{p+q}} J^{m_1 m_2} = 0 \;.
}
For any form of degree $k> 4$, we take the Hodge dual and decompose in similar fashion. In particular, we can apply this procedure to the exterior derivatives of the $SU(4)$-structure itself, leading to
\eq{
\d J &= W_1 \lrcorner \O^* + W_3 + W_4 \wedge J + \text{c.c.} \\
\d \O &= \frac{8 i}{3} W_1 \wedge J \wedge J + W_2 \wedge J + W_5^* \wedge \O \;,
}
with $W_{1,4,5}$  $(1,0)$-forms and $W_{2,3}$ primitive $(2,1)$-forms. The $W_j$ are known as the torsion classes, and are obstructions to integrability of the $SU(4)$-structure on $M_8$. In particular, we have the following table:\footnote{Manifolds with trivial canonical class (i.e. of $SU(4)$-structure) and integrable complex or symplectic structure are sometimes labeled in the literature as ``complex Calabi-Yau" or ``symplectic Calabi-Yau" respectively. We shall not use such terminology;  in our definition, only manifolds with torsion-free $SU(4)$-structure are labeled as Calabi-Yau.}
\begin{center}
\begin{tabular}{|l|l|}
\hline
Geometry of $M_8$ & Torsion classes \\ \hline
Complex & $W_1 = W_2 = 0$ \\ \hline
Symplectic & $W_1 = W_3 = W_4 = 0$ \\ \hline
K\"{a}hler & $W_1 = W_2 = W_3 = W_4 = 0$ \\\hline
Nearly Calabi-Yau & $W_1 = W_3 = W_4 = W_5 = 0$ \\ \hline
Conformal Calabi-Yau & $W_1 = W_2 = W_3 =0$, $2 W_4 =  W_5$ exact \\\hline
Calabi-Yau & $W_j = 0 $ $\forall j$ \\\hline
\end{tabular}
\end{center}
In order to construct the supersymmetric $SU(4)$-structures of the 2A and 2B systems, it will be necessary to know how the torsion classes transform under conformal transformations $g \rightarrow e^{2\o} g$. It follows from  \eqref{eq:su} that $J \rightarrow e^{2 \o} J$, $\O \rightarrow e^{4 \o} \O$, thus the torsion classes transform as
\eq{
W_1 &\rightarrow W_1 \\
W_2 &\rightarrow e^{2 \chi} W_2 \\
W_3 &\rightarrow e^{2 \chi} W_3 \\
W_4 &\rightarrow W_4 + 2 \p^+ \chi \\
W_5 &\rightarrow W_5 + 4 \p^+ \chi\;,
}
with $\p^+ \chi \equiv \pi^{1,0}_I(\d \chi)$.

\section{Useful formulae}
The following formulae were used to compute \eqref{eq:sympsusy}:
\eq{
\star_8 F_1 &= \frac16 i \left(f^{(1,0)}_1 - f_1^{(0,1)} \right)\wedge J \wedge J \wedge J \\
\star_8 F_3 &= \frac12 i f_3^{(1,0)} \wedge J \wedge J + \tilde{f}_3^{(0,1)} \lrcorner \O - i f_3^{(2,1)} \wedge J + \text{c.c.} \\
(\tilde{h}_3^{(1,0)} \lrcorner \O^*) \wedge \O &= - \frac{8}{3} i \tilde{h}_3^{(1,0)}\wedge J \wedge J \wedge J \;.
}

\end{document}